\documentclass[10pt,twocolumn,letterpaper]{article}

\usepackage{wacv}
\usepackage{times}
\usepackage{epsfig}
\usepackage{graphicx}
\usepackage{amsmath}
\usepackage{amssymb}
\usepackage{booktabs}
\usepackage{enumitem} %
\usepackage{wrapfig}

\usepackage{floatrow}
\newfloatcommand{capbtabbox}{table}[][\FBwidth]

\usepackage[utf8]{inputenc} %
\usepackage[T1]{fontenc}    %
\usepackage{url}            %
\usepackage{booktabs}       %
\usepackage{amsfonts}       %
\usepackage{nicefrac}       %
\usepackage{microtype}      %
\usepackage{multirow}

\usepackage[accsupp]{axessibility}  %

\def\wacvPaperID{476} %

\wacvfinalcopy %

\ifwacvfinal
\usepackage[breaklinks=true,bookmarks=false]{hyperref}
\else
\usepackage[pagebackref=true,breaklinks=true,colorlinks,bookmarks=false]{hyperref}
\fi

\newif\ifdraft
\drafttrue

\newcommand{\hr}[1]{{\color{black} #1}}

\newcommand{\HR}[1]{{\color{black}{#1}}}

\newcommand{\cj}[1]{{\color{black}{#1}}}

\newcommand{\ours}{AudioViewer}

\newcommand{\comment}[1]{}

\newcommand{\TODO}[1]{\textcolor{cyan}{TODO: #1}}

\newcommand{\R}{\mathbb{R}}

\newcommand{\parag}[1]{\noindent\textbf{#1}}

\newcommand{\va}{\mathbf{a}}

\newcommand{\vm}{\mathbf{m}}

\newcommand{\vu}{\mathbf{u}}

\newcommand{\vx}{\mathbf{x}}

\newcommand{\vz}{\mathbf{z}}

\newcommand{\mA}{\mathbf{A}}

\newcommand{\mI}{\mathbf{I}}

\newcommand{\mM}{\mathbf{M}}

\newcommand{\mV}{\mathbf{V}}
\newcommand{\mW}{\mathbf{W}}

\begin{document}

\title{AudioViewer: Learning to Visualize Sounds}

\author{Chunjin Song\thanks{Equal contribution}, Yuchi Zhang$^*$, Willis Peng, Parmis Mohaghegh, Bastian Wandt, Helge Rhodin\\
University of British Columbia\\
{\tt\small \{chunjins,wandt,rhodin\}@cs.ubc.ca}
}

\maketitle
\thispagestyle{empty}

\begin{abstract}
A long-standing goal in the field of sensory substitution is enabling sound perception for deaf \hr{and hard of hearing (DHH)} people by visualizing audio content.
Different from existing models that translate \hr{to hand sign language}, between speech and text, or text and images, we target immediate and low-level audio to video translation that applies to generic environment sounds as well as human speech. Since such a substitution is artificial, without labels for supervised learning, 
our core contribution is to build a mapping from audio to video that learns from unpaired examples via high-level constraints.
For speech, we additionally disentangle content from style, such as gender and dialect. 
Qualitative and quantitative results, including a human study, demonstrate that our unpaired translation approach maintains important audio features in the generated video and that 
videos of faces and numbers are well suited for visualizing high-dimensional audio features that can be parsed by humans to match and distinguish between sounds and words. 
Code and models are available at \url{https://chunjinsong.github.io/audioviewer}
\end{abstract}

\section{Introduction}
Humans perceive their environment through diverse channels, including vision and hearing. Because impairment in any sense can lead to drastic consequences, various approaches have been proposed to substitute lost senses, going all the way to recently popularized attempts to directly interface with neurons in the brain (e.g., NeuraLink \cite{musk2019integrated}).
One of the least intrusive approaches is to substitute audio with video, which is, however, challenging due to their high throughput and different modality.

\begin{figure*}[t!]
	\centering
	\includegraphics[width=0.95\textwidth]{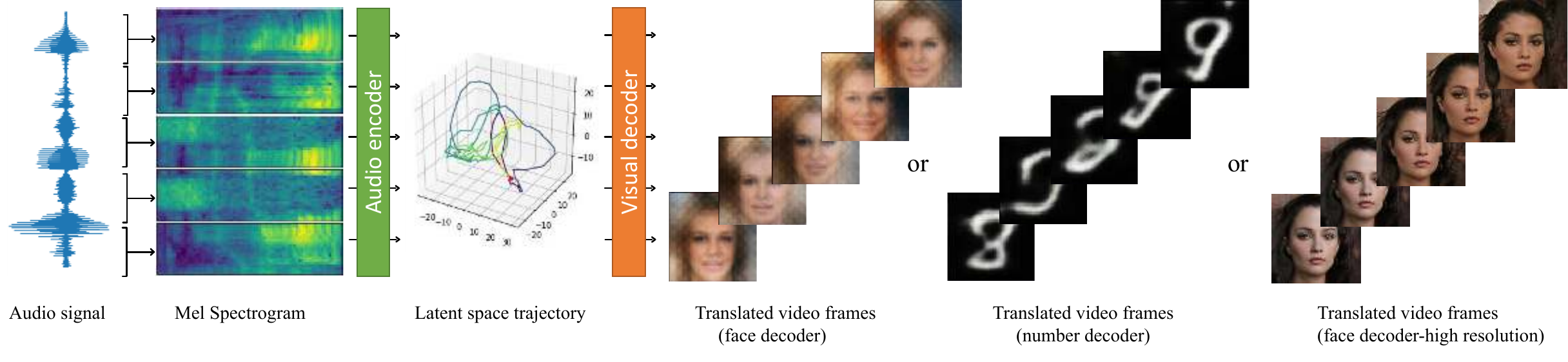}
	\caption{\textbf{\ours:} A tool developed towards the long-term goal of helping hearing impaired persons to see what they can not hear. We map an audio stream to video and thereby use faces or numbers for visualizing the high-dimensional audio features intuitively. Different from the principle of lip reading, this can encode general sound and pass on information on the style of spoken language.}
	\label{fig:teaser}
\end{figure*}

In this paper, we propose a method to visualize audio with natural images in real-time, forming a live video that characterizes the audio content. 
It is a digital sign language with its own throughput, abstraction, automation, and readability trade-offs.
Fig.~\ref{fig:teaser} gives an overview of how short audio segments are sequentially mapped to frames of a live video of transforming figures.

Multiple approaches for audio-to-video translation exist, each striking a different compromise tailored for their application scenario.
Lip-syncing the facial expressions of a virtual avatar from spoken audio \cite{duarte2019wav2pix,sadoughi2019speech,wiles2018x2face,chen2019soundtovisual,mama2021nwt,zhou2020makeittalk} allows deaf and hard of hearing (DHH) people to lip-read spoken texts. However, natural lip motion is a result of speaking and only contains a fraction of the audio content, which we validate in a comparative human study. 
Speech can also be translated to words with a recognition system \cite{noda2015audio}. For example, the spoken 'dog' would be translated to the text 'dog'. %
This is intuitive, but such a translation is still indirect and does not contain any vocal feedback or style differences between male and female speakers~\cite{stewart1976a,zaccagnini1993effects,oster1995teaching}.
Moreover, environmental sounds that cannot be indicated by a single word or lip motion, such as the echo of a dropped object or the repeating beep of an alarm, are ill-represented by all of these existing techniques.
Traditional visual tools for deaf speech learning rely on spectrogram representations \cite{elssmann1987speech,yang2010speech,xu2008speech,kroger2010audiovisual}, which are general but create unnatural images that humans rarely perceive in their surroundings and consequently have difficulties to comprehend. \looseness=-1

\comment{
we seek for an intuitive and immediate visual representation of the high-dimensional audio data and are inspired by early techniques from the 1970s that visualize general high-dimensional data with facial line drawings with varying length of the nose and curvature of the mouth~\cite{chernoff1973use,jacob1976face}.
In this paper, we design an immediate audio to video mapping leveraging ideas from computer vision and audio processing for unpaired domain translation.

Our solution is a direct, low-level translation from audio to video 
via a linked latent space, learned and structured with variational auto encoders (VAEs)~\cite{kingma2013auto} that capture natural statistics and which we equip with additional priors. %
For instance, when mapping from audio to portrait visuals, \TODO{we noticed that speakers of the same sex result in uniform changes in background and hair style across words.} This directness comes at the expense of loosing high-level semantics, such as lip motion, and requires the human user to do a significant amount of learning. A trade-off that we deem worth exploring, with many directions of future work.
}

To overcome these limitations, 
we design \ours{} as a low-level mapping---at the level of sounds and phones instead of words---which makes it immediate
and could be used by infants before they have an understanding of words and language in general. It is similar to visualizing sounds with spectrograms but includes a machine learning component that disentangles factors and makes visualizations easier to comprehend by creating a series of natural images. The difficulty of creating such translation is that there is no canonical association to create paired labels between audio and video\footnote{besides lip motion, which is non-injective and therefore insufficient.}, which is necessary for supervised learning. \looseness=-1

The underlying idea is to find high-level constraints that can be implemented as loss functions for machine learning and
align the representation with human perception. We start from existing unpaired image translation techniques and equip them with additional properties.
Our core contributions to realize perception-driven translation is the design, implementation, and evaluation of the following principles:
\begin{itemize}[noitemsep,topsep=0pt]
\item To maintain the frequency of features, i.e., to map unusual sounds to unusual images, 
we utilize unsupervised learning and exploit cycle consistency to learn a joint structure between audio and video modalities. 
\item Humans are good at recognizing patterns that they see in their environment, particularly faces \cite{aguirre1998area,kanwisher1999fusiform,tarr2000ffa,chernoff1973use,jacob1976face}. So we analyze mapping to %
natural image%
, including faces and hand-written digits of varying resolution.
\item To separate styles, such as gender and dialect, from the content of individual phones and sounds, we disentangle the content by utilizing weak annotations of speech datasets.\looseness=-1
\item Humans are able to perceive complex spatial structures but quick and non-natural (e.g., flickering) changes lead to disruption and tiredness \cite{sperling1960information}. Hence, we enforce smoothness constraints on the learned mapping. \looseness=-1
\end{itemize}

\noindent

We demonstrate the feasibility of this \ours{} approach with a working prototype and quantify the improvement brought about by our contributions by
\begin{itemize}[noitemsep,topsep=0pt]
    \item \cj{developing a low-level mapping from general sound to vision to help DHH people perceive the sound;} 
    \item introducing a new metric on the throughput, a lower bound for the loss of information content;
    \item a human study showing that words and phones can be better distinguished from the generated video features;
    \item a second human study which shows that a set of words in our novel representation can be learned with a success rate~of~87\% in as few as 16 attempts.
\end{itemize}
Clear improvements are gained over synthesized lip motions and plain spectrograms because a partial mapping loses important audio features and because content and style remain entangled in these baselines.

\section{Related Work}

In the following section, we first review the literature on audio and video generation, with a particular focus on cross-modal models. We then put our approach in context with existing assistive systems.

\parag{Classical audio to video translation.} There is a large body of literature on audio-visual learning. We refer interested readers to recent survey~\cite{zhu2021deep} for a more comprehensive overview. Audio to video translation methods have mostly been designed as audio to scenes ~\cite{wan2019towards,qiu2018image,christensen2020batvision, hao2018cmcgan}, audio to motions~\cite{taylor2017deep,karras2017audio,suwajanakorn2017synthesizing,chen2017deepcrossmodal,shlizerman2018audio} and audio to talking faces~\cite{duarte2019wav2pix,sadoughi2019speech,wiles2018x2face,kr2019towards,jamaludin2019you,prajwal2020Lip,zhou2019talking,chen2019soundtovisual,zhou2020makeittalk,mama2021nwt}. However, the translated scene images and body motions generated from sounds are at a high abstraction level, such as mapping the word dog to the image of a dog, but do not contain any vocal feedback, preventing people from learning the sounds from the generated images directly. The most related works are those aiming for digital dubbing or lip-syncing facial expressions to spoken audio~\cite{duarte2019wav2pix,sadoughi2019speech,wiles2018x2face,kr2019towards,jamaludin2019you,prajwal2020Lip,zhou2019talking,chen2019soundtovisual,zhou2020makeittalk,mama2021nwt}. In contrast to our setting, these tasks are typically learned from paired examples, videos with audio lines, e.g., a talking person where the correspondence of lip motion, expressions and facial appearance to the spoken language is used to train the relation between sound and mouth opening. However, there are multiple sounds that correlate with the same lip and facial motion and others that do not correlate at all. By contrast, our unpaired translation mechanism is designed to map the entire audio spectrum. \looseness=-1

\comment{
\parag{Classical audio to video mappings.} Audio to video translation methods have mostly been designed for digital dubbing or lip-syncing facial expressions to spoken audio~\cite{duarte2019wav2pix,sadoughi2019speech,wiles2018x2face,kr2019towards,jamaludin2019you,prajwal2020Lip,zhou2019talking,chen2019soundtovisual,zhou2020makeittalk,mama2021nwt}. Other approaches reconstruct facial attributes, such as gender and ethnicity, from audio and generating matching facial images~\cite{wen2019reconstructing} and map
music to facial or body animation~\cite{taylor2017deep,karras2017audio,suwajanakorn2017synthesizing,chen2017deepcrossmodal,shlizerman2018audio}.
In contrast to our setting, these tasks are typically learned from paired examples, videos with audio lines, e.g., a talking person where the correspondence of lip motion, expressions and facial appearance to the spoken language is used to train the relation between sound and mouth opening. However, there are multiple sounds that correlate with the same lip and facial motion and others that do not correlate at all. By contrast, our unpaired translation mechanism is designed to map the entire audio spectrum. \looseness=-1
}

\parag{Audio and video generation models.}
Image generation models predominantly rely on GAN \cite{goodfellow2014generative} and VAE \cite{kingma2013auto,Hou2017} formulations. The highest image fidelity is attained with hierarchical models that inject noise and latent codes at various network stages by changing their feature statistics \cite{huang2017arbitrary,karras2019style}. 
For audio, only a few methods operate on the raw waveform~\cite{kamper2019truly}. It is more common to use spectrograms and to apply convolutional models inspired by the ones used for image generation \cite{hsu2017learning,dong2018musegan,briot2017deep}. We use cross-modal VAEs as a basis to learn the important audio and image features in terms of their frequency.
 
\parag{Cross-modal latent variable models.}
CycleGAN and its variations~\cite{zhu2017unpaired,taigman2016unsupervised} have considerable success in performing cross-modal unsupervised domain transfer, for medical imaging~\cite{hiasa2018cross,tmenova2019cyclegan} and audio to visual translation~\cite{hao2018cmcgan}, but often encode information as a high-frequency signal that is invisible to the human eye and susceptible to adversarial attacks~\cite{chu2017cyclegan}. An alternative approach involves training a VAE, subject to a cycle-consistency condition ~\cite{jha2018disentangling,yook2020many}, but these works were restricted to domain transfers within a single modality. Most similar is the joint audio and video model proposed by Tian et al.~\cite{tian2019latent}, which uses a VAE to map between two incompatible latent spaces using supervised alignment of attributes. However, it operates on a word, not phoneme level, and has no mechanism to ensure temporal smoothness nor information throughput. Our contributions address these shortcomings. Relatedly, encoder-decoder and GAN models have been applied to generating video reconstructions of lip movements based on audio data~\cite{chung2017you,chen2018lip,vougioukas2018end,zhou2019talking,zhou2020makeittalk,mama2021nwt}, however, due to mapping ambiguities between phonemes and visemes, lip movements are not a reliable source of feedback for learning sound production~\cite{lidestam2006visual,newman2010limitations,cappelletta2012phoneme,fernandez2017optimizing}, which we further confirm with our evaluation.

\parag{Deaf speech support tools.} 
Improvements in speech production for \hr{DHH people} have been achieved through non-auditory aids and these improvements persist beyond learning sessions and extend to words not encountered during the sessions \cite{stewart1976a,zaccagnini1993effects,oster1995teaching}. While electrophysiological \cite{hardcastle1991visual,katz2015visual} and haptic learning aids \cite{eberhardt1993omar} have demonstrated efficacy for improving speech production, such techniques can be more invasive, especially for young children, as compared to visual aids.
Elssmann et al.~\cite{elssmann1987speech} demonstrate visual feedback from the Speech Spectrographic Display (SSD)~\cite{stewart1976a} is equally effective at improving speech production as compared with feedback from a speech-language pathologist. Alternative graphical plots generated from transformed spectral data have been explored by~\cite{yang2010speech,xu2008speech,kroger2010audiovisual}, which aim at improving upon spectrograms by creating plots that are more distinguishable with respect to speech parameters. Other methods aim at providing feedback by explicitly estimating vocal tract shapes~\cite{park1994integrated}. In addition, Levis et al.~\cite{levis2004teaching} demonstrate that distinguishing between discourse-level intonation (intonation in conversation) and sentence-level intonation (intonation of sentences spoken in isolation) is possible through speech visualization and argues that deaf speech learning could be further improved by incorporating the former. Commercially, products such as the IBM's Speech Viewer \cite{ibm2004speech} are available to the public.
Our image generation approach extends these spectrogram visualization techniques by leveraging the generative ability of VAEs in creating a mapping to a more natural 
video representation, which we show leads to improved recognition.  \looseness=-1

\parag{Sensory substitution and audio visualization.}
Related to our work is the field of sensory substitution, whereby information from one modality is provided to an individual through a different modality. While many sensory substitution methods focus on substituting visual information into other senses like tactile or auditory stimulation to help visual rehabilitation \cite{maidenbaum2014sensory,Hu_2019_CVPR,goldish1974optacon}, few methods target substituting the auditory modality with visualization.
Music visualization approaches generate visualizations of songs such that users can browse songs more efficiently without listening to them~\cite{yoshii2008music,takahashi2018instrudive}. On the learning side, \cite{yuan2019speechlens} visualize the intonation and volume of each word in speech by the font size enables learning narration strategies. 
Different from the introduced above, our model tries to visualize speech and other audio at the phoneme and sound level with deep learning models instead of selected hand-crafted features. \looseness=-1

\begin{figure*}[htp]
	\centering
	\includegraphics[width=0.98\textwidth,trim=0 12.4cm 0 0,clip]{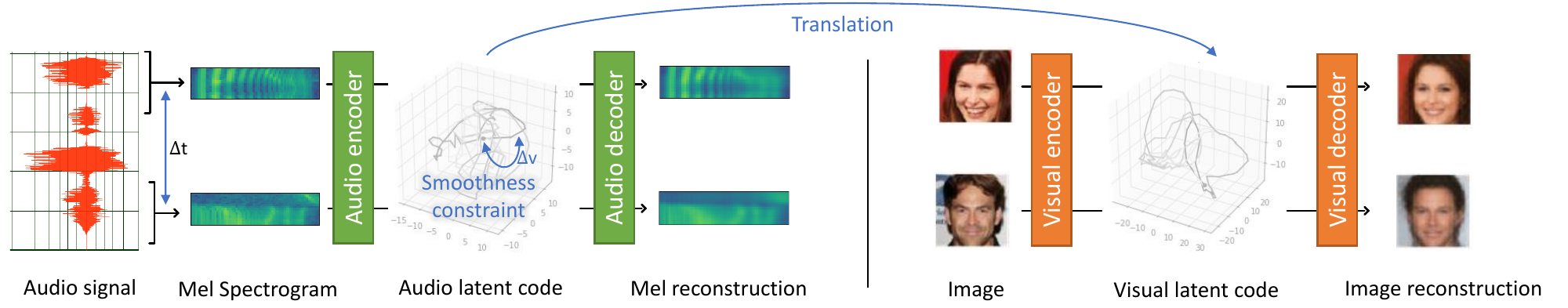}
	\caption{\textbf{Overview.} A joint latent encoding is learned with audio and video VAEs that are linked with a translation network and augmented with a smoothness constraint ($\Delta v$ annotation) and supplemented with disentangled training.}
	\label{fig:overview}
\end{figure*}
\section{Method}%
Our goal is to translate a audio signal $\mA = (\va_1, \cdots, \va_{T_A})$ into a video visualization $\mV = (\mI_1, \cdots, \mI_{T_V})$, where $\va_i \in \R$ are sound wave samples recorded over $T_A$ time steps and $\mI_i$ are images representing the same content over $T_V$ frames. Supervised training is not possible in the absence of paired labels.\footnote{E.g., speech videos contain paired examples, but only for mapping from speech to lip motion, which we show is insufficient.}
Instead, 
we start by learning individual audio and video models without correspondence. These are subsequently linked with an unpaired audio-to-video translation network that conserves high-level properties such as smoothness, regularity, and information loss using cycle-consistency and other unsupervised objectives motivated in the introduction.
Fig.~\ref{fig:overview} shows the individual steps. The audio encoder $E_A(\va_i)$ yields a audio latent code $\vz_i \in \mathcal{Z_A}$ and the visual decoder $D_V(\vz_i)$ outputs a corresponding image $\mI_i$. This produces a video representation of the audio when applied sequentially. Section~\ref{sec:cycle} introduces the handling of mismatching latent spaces with a translation network. \looseness=-1

\parag{Equalizing sound and video dimensionality.} A first technical problem lies in the higher audio sampling frequency (16000 Hz), that prevents a one-to-one mapping to 25 Hz video. We follow common practice and represent the sound wave with a %
mel-scaled spectrogram, $\mM = (\vm_1, \dots, \vm_{T_M})$, $\vm_i \in \R^F$, where $F=80$ is the number of filter banks. 
It is computed via the short-time Fourier transform with a 25 ms Hanning window with 10 ms shifts. 
 In the following, we explain how to map from overlapping segments of length $T_M=20$ (covering $200$ ms) to corresponding video frames.

\subsection{Audio Encoding}
\label{sec:latent models}

Given unlabelled audio and video sequences, we start by learning independent encoder-decoder pairs $(E_A,D_A)$ for sound and $(E_V,D_V)$ for video. We use probabilistic VAEs since these do not only learn a compact representation of the latent structure, but also allow us to control the shape of the latent distribution to be a standard normal distribution.
Let $\vx$ be a sample from the unlabelled audio set. We optimize over all samples using the VAE objective~\cite{kingma2013auto}:
\begin{equation}
\begin{split}
{\mathcal {L}}(\vx ) = -D_{\mathrm {KL} }(q_\phi(\mathbf {\vz} |\vx )\Vert p_{\theta }(\vz )) \\ +\mathbb {E} _{q_{\phi }(\mathbf {\vz} |\vx )}{\big (}\log p_{\theta }(\vx |\vz ){\big )},
\end{split}
\end{equation}
with $D_{\mathrm {KL} }$, the Kullback-Leibler divergence, $q_{\phi }(\vz |\vx )$ and $p_{\theta }(\vx |\vz )$, the latent code and output domain posterior, respectively. These have a parametric form, with%
\vspace{-10pt}%
\begin{align}%
q_{\phi }(\vz |\vx ) &={\mathcal {N}}({\boldsymbol {\rho }}(\vx ),{\boldsymbol {\omega }}^{2}(\vx )\mathbf {I} )\text{ and }\\
p_{\theta }(\vx |\vz ) &={\mathcal {N}}({\boldsymbol {\mu }}(\vz ),{\boldsymbol {\sigma }}^{2}(\vz )\mathbf {I} ),
\end{align}
where ${\boldsymbol {\rho }}$ and $\boldsymbol {\omega }$ are the output of the encoder and $\boldsymbol {\mu }$ and $\boldsymbol {\sigma }$ the output of the decoder. \cj{The dimention of $z$ is 256.}
We use the SpeechVAE model from Hsu et al.~\cite{hsu2017learning} that is widely used in sound generation. \looseness=-1

\comment{
\parag{Audio network architecture.}
\label{sec:sound_VAE}

We use the SpeechVAE model from Hsu et al.~\cite{hsu2017learning} that is widely used for generative sound models. Fig.~\ref{fig:overview}, left, sketches how the mel spectrogram is encoded with an encoder $E_A$ of three convolutional layers followed by a fully connected layer which flattens the spatial dimensions. The mel spectrogram $\mM$ is a two-dimensional array, spanning the time and frequency, respectively. To account for the structural differences of the two, convolutions are split into separable $1\times F$ and $3\times 1$ filters, where $F$ is the number of frequencies captured by $\mM$ and striding is applied only on the temporal axis. The decoder is symmetric to the encoder. We use ReLU activation and batch normalization layers in the encoder and decoder.}

\subsection{Structuring the Audio Encoding}
\label{sec:priors}

\paragraph{Content separation.} To better model speech, which we expect to have a higher information content compared to environment sounds, we further disentangle the style, such as gender and dialect, from the content conveyed in phones by leveraging datasets that have phone and speaker annotations. The separation follows that of established audio models \cite{qian2019autovc}. \cj{Here we constrain the disentanglement by a recombined reconstruction loss term, $L_{rr}$}. The details are provided in the supplemental file. Unless otherwise stated, we map only the audio content for visualization in our experiments.

\parag{Smoothness.} We desire our latent space to change smoothly in time. 
However, the audio encoder has a small temporal receptive field, encoding time segments of 200ms. This lets encodings of subsequent sounds be encoded to distant latent codes leading to quick visual changes in the decoding. 
To counteract, we add a temporal smoothness loss on pairs of mel spectrogram segments $\{\mathbf{M}_{i,1},\mathbf{M}_{i,2}\}$ sampled at random time steps $\{t_{i,1} ,t_{i,2}\}$ spaced at most 800ms apart. We test two different pair-loss functions to enforce temporal smoothness in the embedded content vectors. First, by making changes in the latent space proportional to changes in time,\looseness=-1
\begin{equation}
\mathcal{L}_{p, MSE} = \textstyle
\frac{1}{N}\sum_{i}^{N}\left(\Delta \hat{\vz}_i- \|t_{i,1}-t_{i,2}\|\right)^2,
\end{equation}
$\Delta \hat{\vz}_i = s_p \cdot\|\mathbf{z}_{i,1}-\mathbf{z}_{i,2}\|$ the distance in the latent space obtained from $\{\mathbf{M}_{i,1},\mathbf{M}_{i,2}\}$, scale $s_p \in \R$ learned to find a scale between time and latent space dimensions. 
We found it advantageous to measure distances in the logarithmic scale,
\begin{equation}
\mathcal{L}_{p, \log} =\textstyle \frac{1}{N}\sum_{i}^{N}\left(\log{\Delta \hat{\vz}_i}- \log{\|t_{i,1}-t_{i,2}\|}\right)^2,
\end{equation}
which lessens the weight for more distant encodings. 
Our final model uses $\mathcal{L}_{p, \log}$
We add this loss to the VAE objective with a weight of $\lambda_p=10^3$.

\subsection{Image Autoencoder for Video Generation}
\label{sec:video_VAE}

We experiment with three different video models ranging from low-res to photo-realistic image generation:
A linear PCA space, the image DFC-VAE model~\cite{Hou2017}, and the image Soft-IntroVAE model~\cite{daniel2021soft}. 
We use pre-trained models, trained on collections of images. At first, naive unpaired translation is attained by encoding audio snippets sequentially and concatenating the audio encoder and video decoder of two VAEs with matching latent dimension and the same prior distribution $p(\vz)$ over $\vz \in \mathcal{Z}_V$ to generate the frames of the output video. \looseness=-1

\subsection{Linking Audio and Visual Spaces}
\label{sec:cycle}

One of our key contributions is to meaningfully link the audio and video domain without paired examples. 
The naive concatenation explained in the previous section leads to low-quality results because smoothness in the latent space does not necessarily lead to smoothness in the output video, latent dimensions often do not match, and both encoders are only approximations to the true distribution. 

\begin{figure*}[t]
	\centering
	\includegraphics[width=0.98\textwidth,trim=0 13.2cm 5.0cm 0,clip]{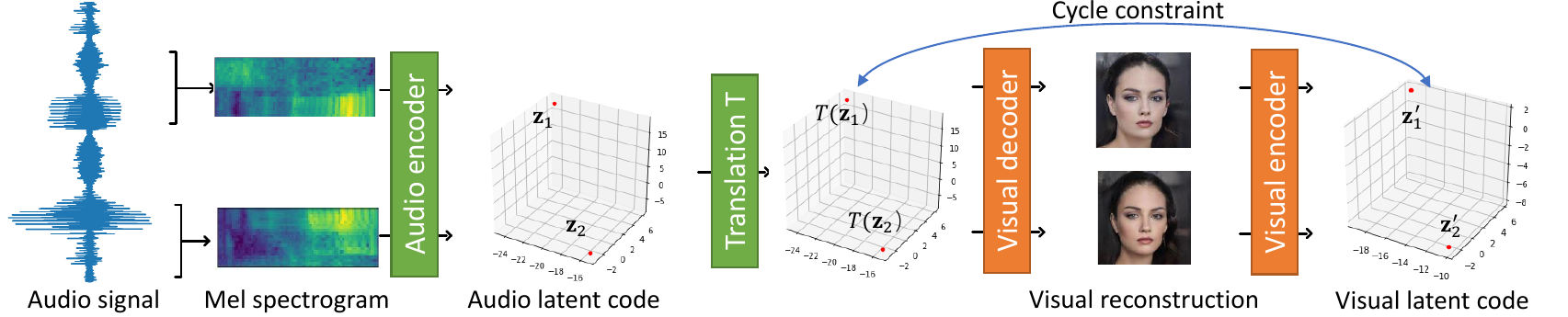}
	\caption{\textbf{Cycle constraint.} We apply a cycle constraint to ensures that the signal is preserved through video decoding and encoding.}
	\label{fig:cycle}
\end{figure*}

When latent spaces have the same dimension, i.e., with \cite{Hou2017} as visual model, we utilize a shared latent space and refine the weights of the video model to bridge structural differences. For \cite{daniel2021soft}, which has a larger latent space and is costly to train, we use a pre-trained model. To nevertheless link audio and video latent spaces, we introduce a translation network $T$ that maps the latent code $\vz$ of the input audio segment to the visual latent variable $T(\vz)$. Fig.~\ref{fig:cycle}, visualizes this mapping and how the visual decoder $D_V$ subsequently decodes $T(\vz)$ to the output image~$I$.

Since our setting is unpaired we have to resort to self-supervised losses for modeling high-level constraints on auxiliary tasks instead of supervised learning. We explain them in the order of basic to more complex. First, to be a proper mapping, the range of $T(\vz)$ should be in the domain of the video decoder. Since the visual latent space fulfills the standard Gaussian distribution, we minimize its negative log-likelihood, \looseness=-1
\begin{equation}
\mathcal{L}_{reg} =\lambda_{reg} \| T(\vz) \|^2,
\end{equation}
where $\lambda_{reg}=1.0$. 
Second, we minimize a lower bound on the information loss with the cycle consistency loss,
\begin{equation}\label{eq_cycle}
\mathcal{L}_{cycle} =\lambda_{cycle} \| E_V(D_V(T(\vz)))-T(\vz) \|,
\end{equation}
that measures the difference between $T(\vz)$ and the reconstructed encoding of the generated image $I$ via visual encoder $E_V$ and is weighted by
$\lambda_{cycle}=1.0$. 

Meanwhile, consistent with Section~\ref{sec:priors}, the changes in the generated image should be smooth. We ensure this with an additional temporal smoothness loss
\begin{equation}\label{eq_smooth}
\begin{split}
\mathcal{L}_\text{smooth} = \lambda_\text{smooth} \frac{1}{N}\sum _i^N(\log(\left \|(I_{i,1} - I_{i,2}) \right \| \\s_\text{smooth})
- \log\left \|t_{i,1} - t_{i,2}\right \|)^2, 
\end{split}
\end{equation}
\cj{which is applied on each pixel of the generated images.}
To further facilitate training of our multi-stage architecture, we also enforce a distance preserving loss \cj{on the latent spaces,}
\begin{equation}\label{eq_distance}
\begin{split}
\mathcal{L}_\text{iso} = \lambda_\text{iso} \frac{1}{N}\sum _i^N(\log\left \|T(\vz_{i,1}) - T(\vz_{i,2})\right \| \\
- \log(\left \|(\vz_{i,1} - \vz_{i,2}) \right \|s_\text{iso}))^2.
\end{split}
\end{equation}

As in Section~\ref{sec:priors}, we input the mel spectrogram segment pairs with random time steps $\{t_{i,1} ,t_{i,2}\}$, where $i$ represents the $i$th input pair. And $\lambda_\text{iso}=1.0$,  $s_\text{iso}=5.0$, $\lambda_\text{smooth}=5.0$ and $s_\text{smooth}=0.001$.
Both Eq.~\ref{eq_smooth} and Eq.~\ref{eq_distance} 
help maintain the feature distances between latent spaces, preventing possible degenerate solutions for the translation network $T$ caused by Eq.~\ref{eq_cycle} effectively.
Note that neither of these self-supervised losses requires annotation. It therefore applies to environment sounds and could easily be finetuned for any language or dialect as long as an audio recording is available.

To measure the throughput of the entire system when using $T$ together with \cite{daniel2021soft}, we train a back-translation network $T^\dagger$ to reconstruct the input audio with
\begin{equation}
    \mathcal{L}_\text{back} = \|D_A(T^\dagger(E_V(D_V(T(E_A(\va_i))))))-\va_i\|^2.
\end{equation}
This reverse map training is a postprocess only used for evaluating throughput.

\comment{
An additional cycle constraint that works without image correspondence and ensures that samples from the audio latent space posterior are reconstructed well, similar in spirit to \cite{jha2018disentangling,yook2020many}. Fig.~\ref{fig:cycle} shows the cyclic chaining from audio latent code to video and back. It is implemented on the content space as
\begin{equation}
\mathcal{L}_{cycle} = \left| E_V(D_V(Map(z)))-z \right|,
\end{equation}
where $z$ is drawn from the posterior of the content part $E_A(\mM)$ of the audio encoder and the video modules operate on mean values minimizing point-wise differences in position. 
We add this constraint when training the audio encoder with weight $\lambda_\text{cycle} =10$.\HR{This section is a bit out of place. Is it really used at training time of the audio encoder. Then we should mention it further up...} 
}

\section{Experiments}
We show qualitatively and quantitatively that \ours{} conveys important audio features via visualizations of portraits or numbers, that it applies to speech and environment sounds, that it outperforms existing baselines, and that the task cannot be solved with mapping to lip motion~\cite{zhou2020makeittalk}.
The supplemental documents provides additional results, including example videos and human study details. \looseness=-1

\parag{Quantitative metrics.}  We compare latent embeddings by the Euclidean distance, quantify differences in the mel spectrogram with the signal to noise ratio (SNR),
 and measure smoothness as the change in latent space position over time.

\parag{Perceptual studies.} We perform two human studies to analyze the human capability to perceive the translated audio. As perceptual results may be subjective, we report statistics over 29 questions answered by 10-22 participants for each of the examined approaches for the distinguishability study and 9 participants for the learnability study. The details are given in the supplemental document.

\parag{Baselines.} We compare to the most straight-forward speech to video translation, by \emph{i)} mapping to lip and facial motion using the recent MakeItTalk method~\cite{zhou2020makeittalk} and \emph{ii)} comparing to spectogram visualizations as used in current assistive systems. To show that simpler approaches are insufficient, we experiment \emph{iii)} with simpler variants of our approach and \emph{iv)} principal component analysis (PCA) as further baselines.
PCA yields a projection matrix $\mW$ that rotates the training samples $\vu_i$ by $\vz_i = \mW \vu_i$ to have 
maximal variance. 
We use the reduced PCA version, where $\mW$ maps to a $d$-dimensional space. %
A further important factor is the impact of the video domain, for which we compare digits vs.~faces at the same resolution and low vs.~high-resolution faces. In addition, we ablate the impact of our latent space priors and training strategies introduced in sections~\ref{sec:priors} and \ref{sec:cycle} as well as when disabling the style-content disentangling.

\parag{Datasets.} We use the TIMIT dataset~\cite{garofolo1993timit} for learning speech embeddings. It contains 5.4 hours of audio recordings (16 bit, 16 kHz) as well as time-aligned orthographic, phonetic and word transcriptions for 630 speakers of eight major dialects of American English, each reading ten phonetically rich sentences. We use the training split (462/50/24 non-overlapping speakers) of the KALDI toolkit~\cite{povey2011kaldi}. The phonetic annotation is only used at audio encoder training time and as ground truth for the human study. 
In addition, we report how well a model trained on speech generalizes to environmental sounds on the ESC-50 dataset \cite{piczak2015esc}.

To test what kind of visualization is best for humans to perceive the translated audio, we train and test our model on three image datasets: The face attributes dataset CelebA-HQ~\cite{karras2017progressive}(29000/1000 images for train/val respectively) and CelebA~\cite{liu2015faceattributes}  (162770/19962 images for train/val respectively), the MNIST~\cite{lecun1998gradient} datasets (60000/10000 images for train/val respectively). All shown results are with the high-res \cite{daniel2021soft} decoder unless otherwise mentioned.%

\parag{Runtime.}
\comment{Speech models without disentanglement are trained for 300 epochs (120,000 iterations) with a batch size of 64. Disentanglement speech models are trained for 600 epochs (165,000 iterations) with a batch size of 64. Note that the total size of triplet dataset is different to the paired dataset due to the requirement of parallel data. CelebA models for 38 epochs and MNIST models for 24 epochs with batch size 144, and joint models are fine-tuned jointly on audio and image examples for 10 audio epochs.} 
The supplemental video contains a live demo. The \ours{} using \cite{Hou2017} is real-time capable, the inference time for a frame is 5 ms (low-res models) / 7.6 ms (high-res models)  on an i7-9700KF CPU at 3.60GHz with a single NVIDIA GeForce RTX 2080 Ti. 

\begin{figure}[b]
	\centering
	\includegraphics[width=0.98\linewidth, trim=0 0 0 0cm,clip]{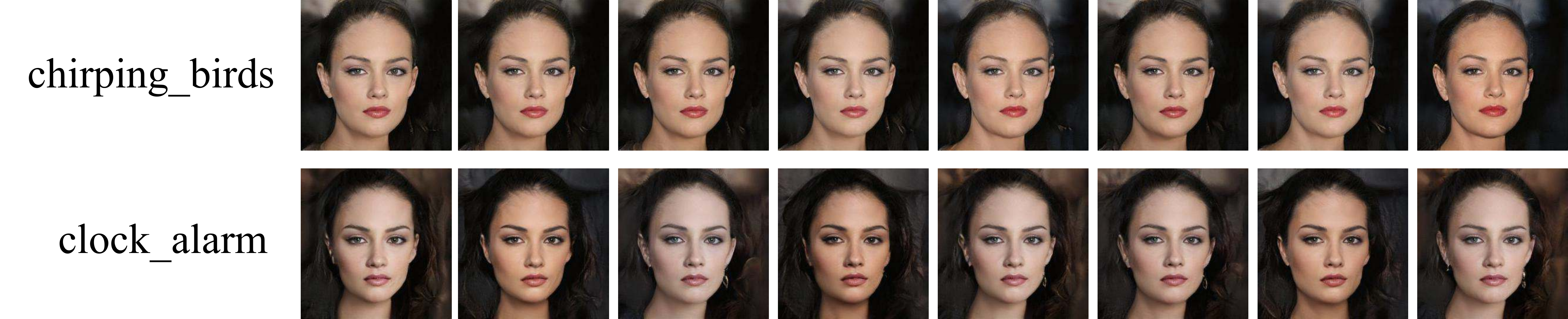}
	\caption{\textbf{Environment sounds visualization.} Although not trained for it, our approach also visualizes natural sounds. Here showing a clear difference between a chirping bird and an alarm clock.}
	\label{fig:environment}
\end{figure}

\comment{
\begin{wrapfigure}{R}{0.48\textwidth}
\centering
\includegraphics[width=0.98\linewidth]{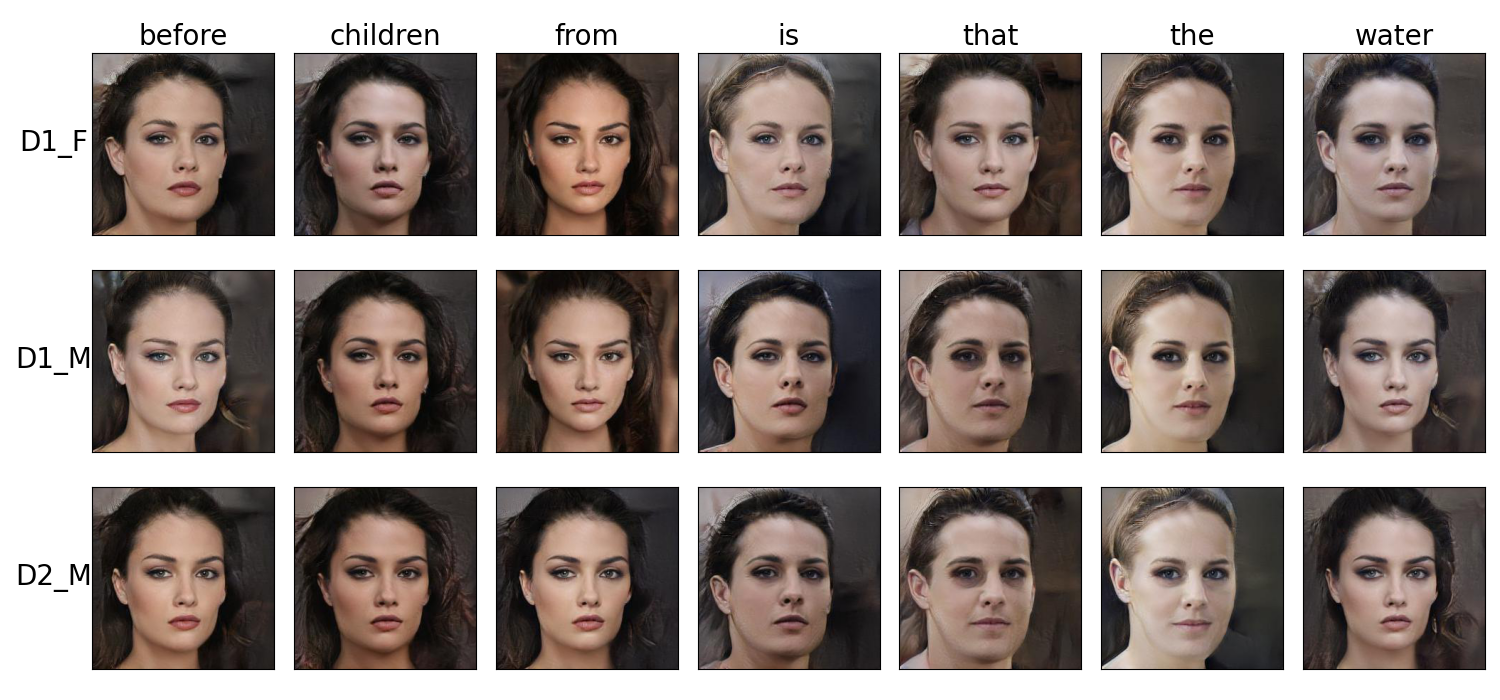}
	\includegraphics[width=0.98\linewidth, trim= 0cm 0cm 0cm 2.0cm,clip]{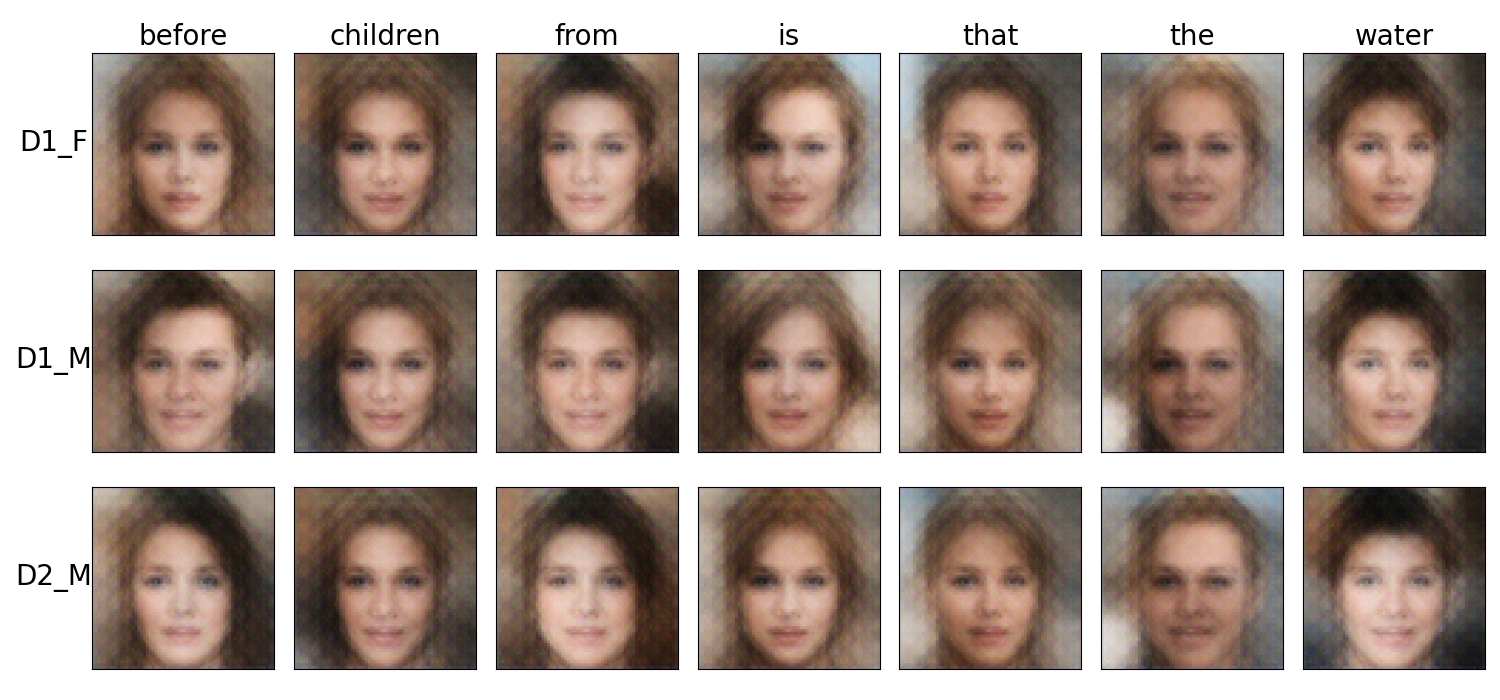}
	\includegraphics[width=0.98\linewidth, trim= 0cm 0cm 0cm 2.0cm,clip]{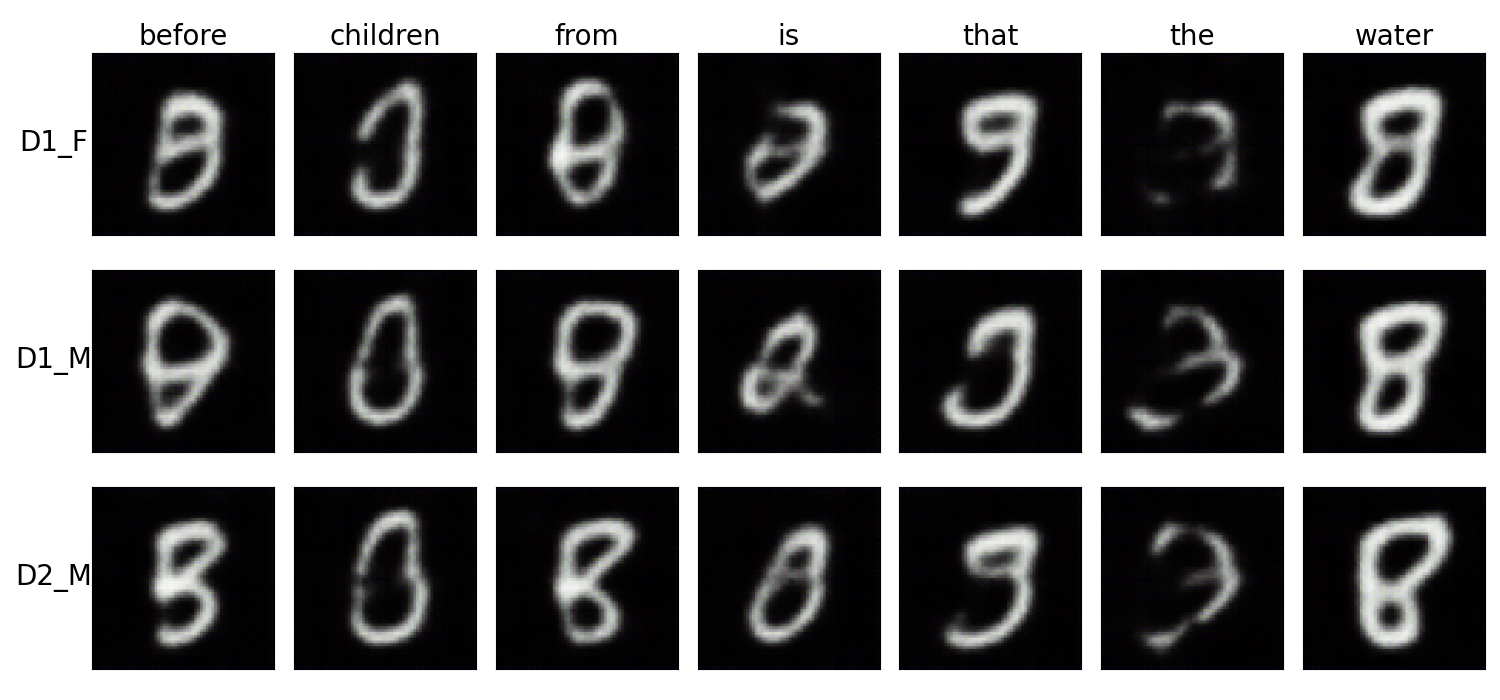}%
\caption{\label{fig:phonemes content} \textbf{Phoneme similarity.} The first phone of different words are distinctive (columns) while the speakers has a small influence (rows). This is consistent across image domains (Top to bottom: CelebA high-res, CelebA low-res, and MNIST).}
\end{wrapfigure}
}

\comment{
\setlength\figlimitscale{0.5\textwidth}
\setlength{\intextsep}{0pt}%
\setlength\figlimitimscale{1.0\figlimitscale}

\begin{wrapfigure}{r}{0.5\textwidth}
\centering
\setlength{\fboxrule}{0pt}%
\parbox[t]{\figlimitimscale}{%
\centering%
\fbox{\includegraphics[width=0.95\linewidth]{images/new_words_content_face_HR.png}
	\includegraphics[width=0.95\linewidth, trim= 0cm 0cm 0cm 2.0cm,clip]{images/new_words_content_face_cycle_log_flipped.png}
	\includegraphics[width=0.95\linewidth, trim= 0cm 0cm 0cm 2.0cm,clip]{images/new_words_content_MNIST_cycle_log_flipped.png}%
}\\%
}%
\caption{\small \textbf{Phoneme similarity.} The first phone of different words are distinctive (columns) while the speakers has a small influence (rows). This is consistent across image domains (Top to bottom: CelebA high-res, CelebA low-res, and MNIST).}
\label{fig:phonemes content}
\end{wrapfigure}
}

\begin{figure}[t]
	\centering
	\includegraphics[width=0.75\linewidth, trim= 0cm 0cm 0cm 0.5cm,clip]{images/new_words_content_face_HR.png}
	\includegraphics[width=0.75\linewidth, trim= 0cm 0cm 0cm 2.5cm,clip]{images/new_words_content_face_cycle_log_flipped.png}
	\includegraphics[width=0.75\linewidth, trim= 0cm 0cm 0cm 2.5cm,clip]{images/new_words_content_MNIST_cycle_log_flipped.png}
	\caption{\textbf{Phoneme similarity.} The first phone of different words are distinctive (columns) while the speakers has a small influence (rows). This is consistent across image domains (Top to bottom: CelebA high-res, CelebA low-res, and MNIST).} %
	\label{fig:phonemes content}
\end{figure}

\comment{
\begin{figure}[t]
	\centering
	\includegraphics[width=\linewidth]{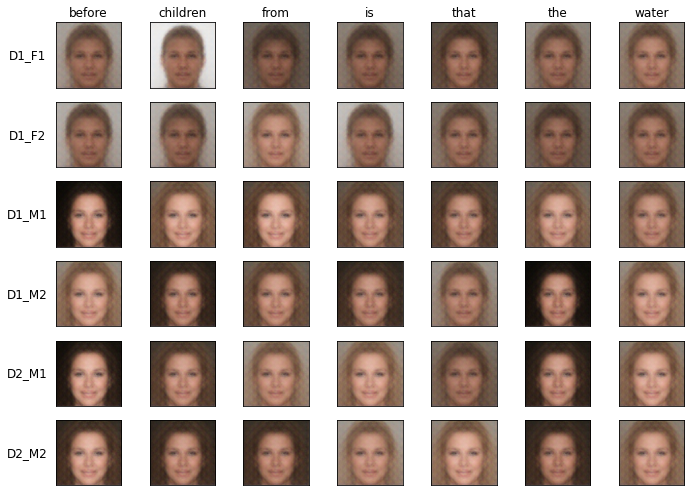}
	\caption{\textbf{Style analysis.} Opposed to the content encoding, the style encoding enhances different speakers influence (rows) while having similar encodings across different words (columns).} %
	\label{fig:phonemes stye}
\end{figure}
}

\begin{figure}[t]
	\centering
	\includegraphics[width=0.9\linewidth]{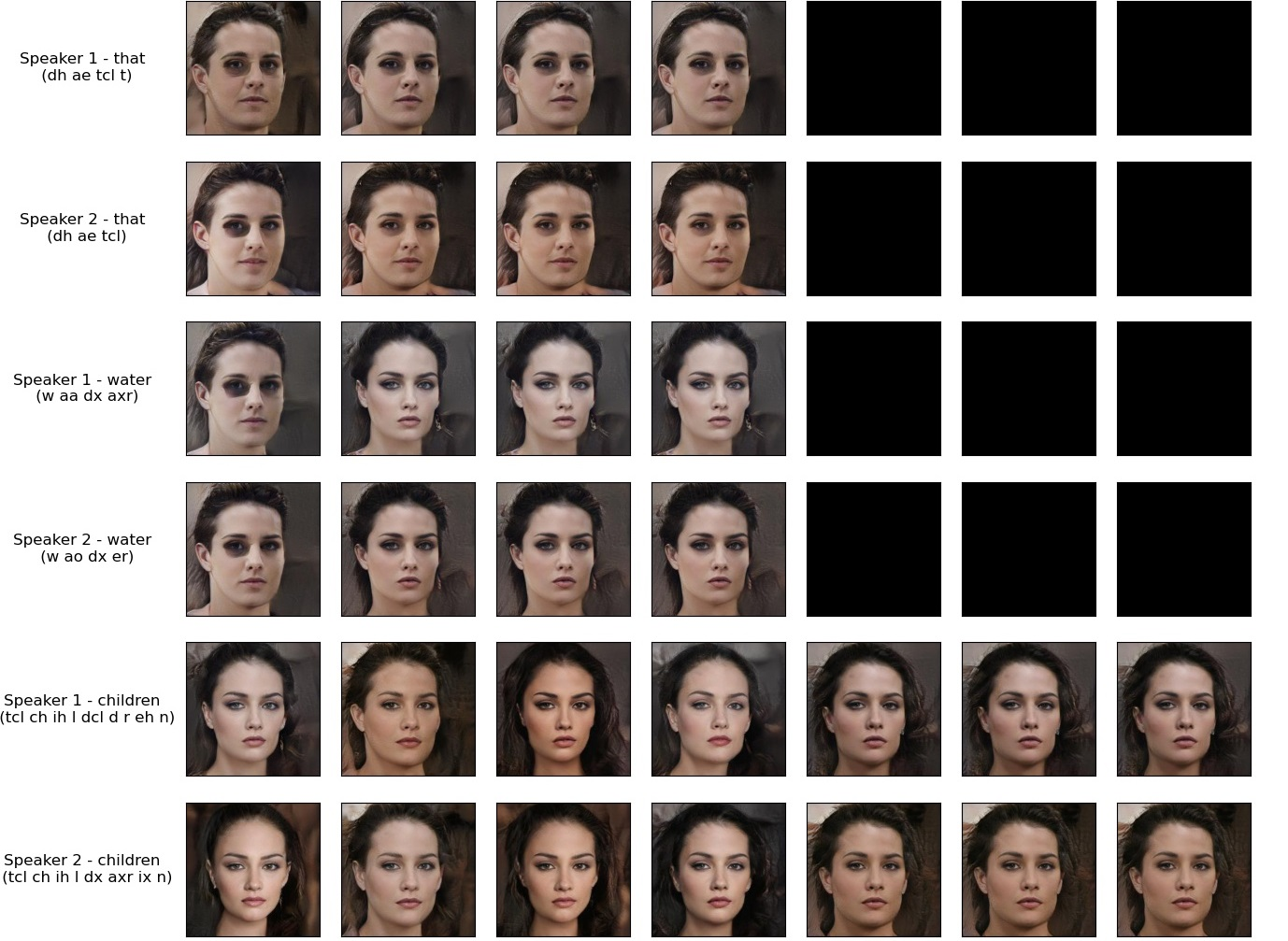}
	\caption{\textbf{Word analysis.} Instances (different rows) of the same words are encoded with a similar video sequence (single row) even though spoken by different speakers.}
	\label{fig:words}
\end{figure}

\comment{
\newlength\figlimitscale
\setlength\figlimitscale{0.5\textwidth}
\setlength{\intextsep}{0pt}%
\newlength\figlimitimscale
\setlength\figlimitimscale{1.0\figlimitscale}
\begin{wrapfigure}{r}{\figlimitscale}
\centering
\setlength{\fboxrule}{0pt}%
\parbox[t]{\figlimitimscale}{%
\centering%
\fbox{\includegraphics[width=0.98\linewidth]{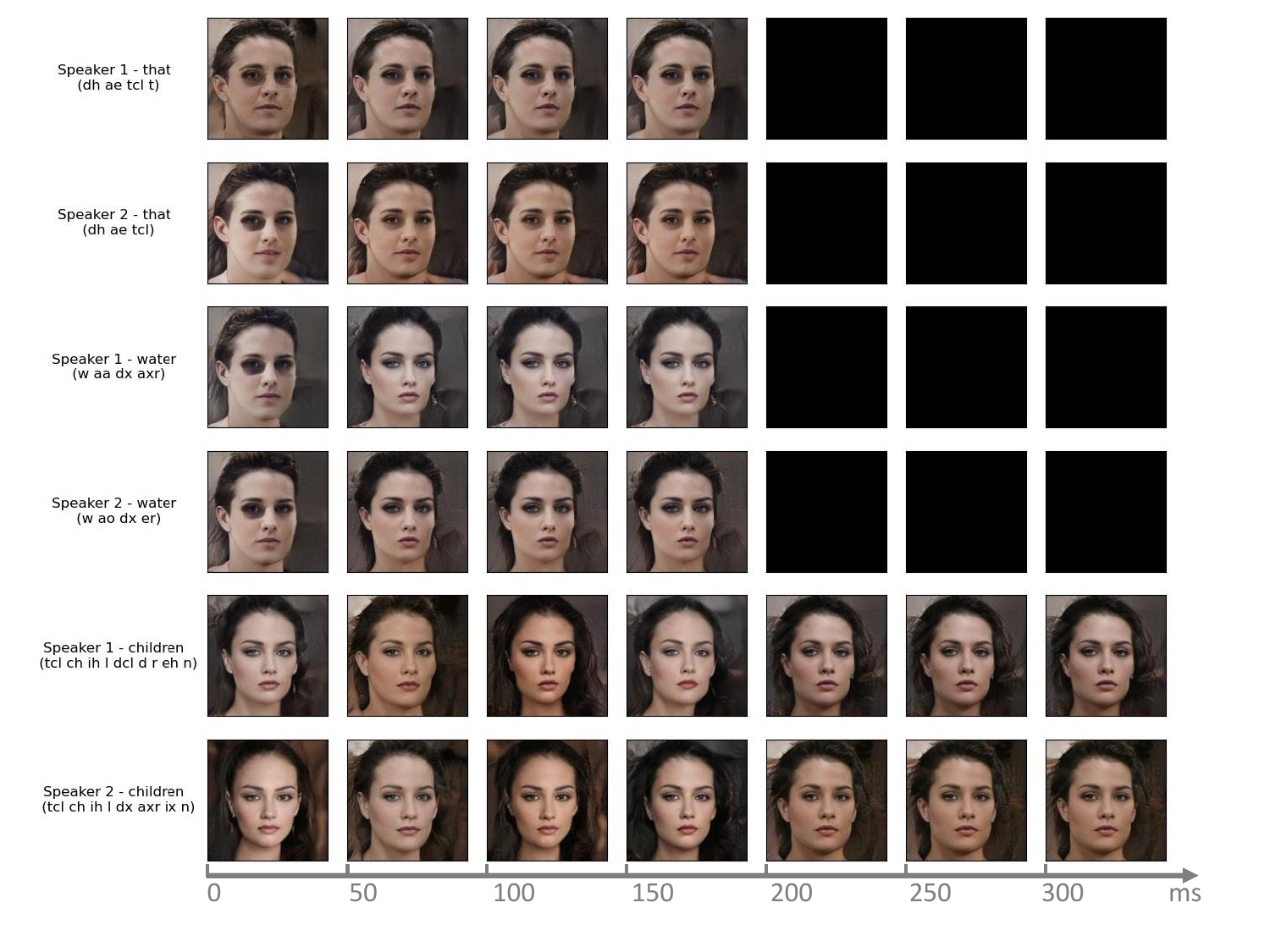}%
}\\%
}%
\caption{ \textbf{Word analysis.} Instances (different rows) of the same words are encoded with a similar video sequence (single row) even though spoken by different speakers.}
\label{fig:words}
\end{wrapfigure}
}

\subsection{Visual Quality and Phonetics} 
A meaningful audio translation should encode similar sounds with similar visuals. Fig.~\ref{fig:phonemes content} depicts such similarity and dissimilarity on the first 200 ms of a spoken word since these can be visualized with a single video frame. This figure compares the content encoding to MNIST and faces; both are well suited to distinguish sounds.
The same word (column) has high visual similarity across different speakers (F\#: female, M\#: male; \# the speaker ID) and dialects (D\#) while words starting with different sounds are visually different. Multiple frames of entire words are shown in Fig.~\ref{fig:words}.
Although trained entirely on speech, our sound-level formulation enables AudioViewer to also visualize environment sounds. Fig.~\ref{fig:environment} gives two examples on bird and alarm sounds. The supplemental contains additional ones.

We compare AudioViewer to the most related methods, including the lip synchronisation method \cite{zhou2020makeittalk} and mel spectogram visualizations as used in current assistive systems, as well as to simpler baselines.
This first study test how well visualizations can be distinguished without requiring a time-intensive training period. This enables us to ask  29 questions to 60 participants and to cover 34 different words and phones, thereby showing generality.

\subsection{Human Study I - Discriminating Sounds}
\begin{figure}[tp]
    \centering
    \includegraphics[width=0.98\textwidth, trim=0cm 0.5cm 2cm 0cm,clip]{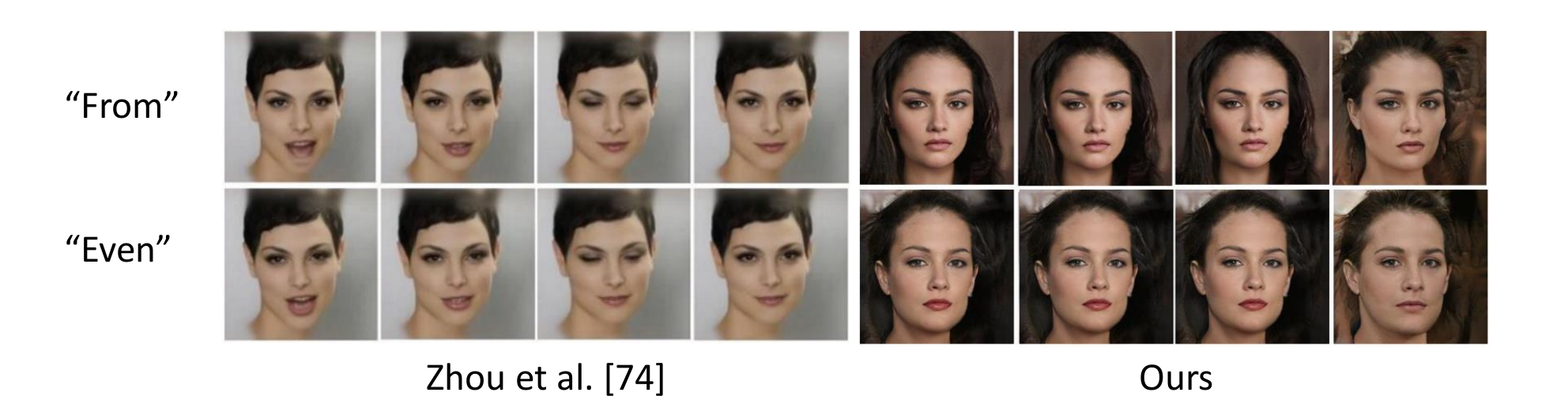}
    \caption{\textbf{Comparison to lip-sync by \cite{zhou2020makeittalk}}. While the lip synchronization looks very similar, the two words are clearly distinguishable in the \ours{} visualizations.}
    \label{fig:makeittalk_comparison}
\end{figure}
\paragraph{Identifying and distinguishing sounds.}
Our human study analyzes the ability to distinguish between visualizations of different sounds, similar to the ones show in Fig.~\ref{fig:phonemes content} and the results are shown in Table~\ref{tab:user_study}. Overall, the users were able to correctly recognize visualizations for the same sound with an accuracy of 86.4\% for the high-res CelebA content model and 85.0\% vs. 79.7\% for the low-res CelebA and MNIST content models, respectively. Broken down by tasks, users for the CelebA model achieve a better accuracy of 87.9\% in correctly matching one of two visualizations against a reference (random guessing yields 50\%), while for the task of grouping four visualizations into pairs, users achieved an accuracy of 85.7 \% (random guessing yields 33.3\%). 
The accuracy scores of MEL are far lower than those of our model, indicating that the patterns in mel spectrogram are more difficult to distinguish than the features generated by our learned model.
In comparison to the lip synchronisation of Zhou et al. \cite{zhou2020makeittalk} \ours{} enables a user to distinguish sounds significantly better. This is due to the almost identical lip motion for different phonemes (e.g. "d", "t" and "n") which makes them indistinguishable.
Fig.~\ref{fig:makeittalk_comparison} shows such a case of two words that look similar in lip motion but are clearly distinguishable in the AudioViewer visualizations.

\parag{Latent space disentanglement.}
Visualizing the style and content part separately with our full model significantly improves recognition scores. The disentangled face model increases the accuracy for distinguishing between speakers of a different sex from $43.3 \pm 2.6\%$ to $ 78.0 \pm 2.9\%$ and speakers of different dialects from $39.6 \pm 7.2\%$ to $56.7 \pm 5.7\%$, with $\pm$ reporting the standard error.
In practice, \ours{} should therefore use the full model and show separate visual decodings side by side or have the option to switch between content and style visualization.
The entire study is reported in the supplemental document. 
\begin{table}[b]
    \caption{\textbf{User Study I results.} User answer accuracy in percent ($\pm$~standard error) for distinguishing between visualizations of different sounds, broken down by question type and phone-pairs vs.~words. \looseness=-1}
    \centering
    \resizebox{0.98\linewidth}{!}{  
    \begin{tabular}{llllll}
    \toprule
Dataset    & MEL & Zhou \cite{zhou2020makeittalk} & MNIST & CelebA & CelebA\\
Resolution    & (spectrum) & (high-res) & (low-res) & (low-res)  & (high-res) \\
    \midrule
        Matching & $65.5 \pm 1.2$  & $65.6 \pm 3.6$  & $77.4 \pm 2.8$ & $84.5 \pm  3.7$& $\textbf{87.9} \pm  \textbf{2.2}$\\
        Grouping & $66.5 \pm 2.1$  & $39.5 \pm 3.9$  & $80.8 \pm 2.2$ & $85.2 \pm 2.4$ & $\textbf{85.7} \pm \textbf{1.8}$ \\
    \midrule
        Phones & $76.5 \pm 1.9$ & $69.5 \pm 5.2$  & $\textbf{91.8} \pm \textbf{2.5}$ & $85.7 \pm 3.0$ & $87.0 \pm 2.5$  \\
        Words &  $60.0 \pm 2.2$  & $39.4 \pm 2.9$  & $73.8 \pm 1.9$ & $84.5 \pm 2.3$ & $\textbf{86.0} \pm \textbf{1.5}$\\
    \midrule 
    \textbf{Overall} &  $66.2 \pm 1.3$  & $47.7 \pm 2.7$ & $79.7 \pm 1.7$ & $85.0 \pm 1.8$ & $\textbf{86.4} \pm \textbf{1.4}$\\
    \bottomrule
    \end{tabular}}
    \label{tab:user_study}
\end{table}
\comment{
\begin{figure}[t]
\floatbox[{\capbeside\thisfloatsetup{capbesideposition={right,top},capbesidewidth=0.38\textwidth}}]{figure}[\FBwidth]
{\caption{\textbf{User Study II results.} Subjects learn over time to distinguish words. The learning speed and accuracy of our model (green) outperform the baseline (red). Accuracy is plotted as the moving average of the success rate over trials.}\label{fig:learning curves}}
{\includegraphics[width=0.60\textwidth, trim=4.5cm 0cm 4.5cm 0.5cm,clip]{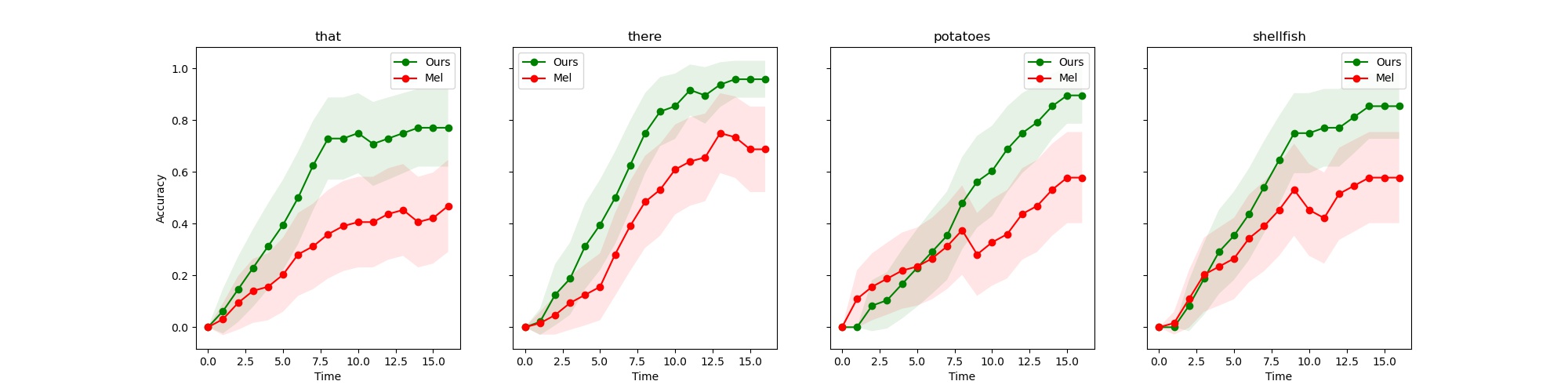}}

\end{figure}
}

\begin{figure}[t]
	\centering
	\includegraphics[width=0.98\linewidth,trim=2cm 0.2cm 1.5cm 0.8cm,clip]{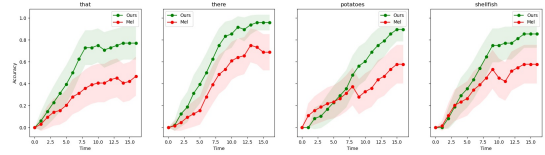}
	\caption{\textbf{User Study II results.} Subjects learn over time to distinguish words. The learning speed and accuracy of our model (green) outperform the baseline (red). Accuracy is plotted as the moving average of the success rate over trials.}
	\label{fig:learning curves}
\end{figure}

\subsection{Human Study II - Learning Sounds}
In this second study we analyze learnability of our best model, determined in Study I and the ablation study, compared to using the best baseline, the mel spectogram (MEL). \cj{Four words were selected that include pairs that sound similar (\emph{that}, \emph{there}) and pairs that have the same length (\emph{potatoes}, \emph{shellfish}).} They are spoken 4 times by different speakers with varying dialects and the same gender.
During the study one word is visualized at a time. The participant is tasked to select the corresponding word. After each selection, the correct answer is given as feedback to enable learning.

Fig.~\ref{fig:learning curves} shows the average learning curves of nine participants that performed the same study for \ours{} (green) and MEL (red). The curves reflect how the participants learn to recognize the same sound spoken by different people over time. The comparison shows that our method performs better than MEL in both learning speed and final accuracy for all of tested words (the shaded area is the standard error). The final accuracy after 16 rounds of learning is Ours 87.0\% vs. MEL 57.8\%, which reflects the learnability of our results. Additional details are included in the supplemental document. \looseness=-1

\comment{In this second study, four phones/words spoken by different speakers with different dialects and the same gender were shown sequentially to the subject. The study task was to select the visualized sound/word from the four choices, e.g, phones \emph{k}, \emph{k\_s}, \emph{tcl}, \emph{d\_aa}. After each selection, the correct answer is disclosed as a feedback mechanism. Fig.~\ref{fig:learning curves} shows the learning curves for one subject with respect to each sound/word. Their increasing trajectory reflects how the subject learns to recognize the same sound spoken by different people over time. The entire study and study details are included in the supplemental document.}
\comment{
\parag{Structural similarities.}
We analyzed whether structural similarities in audio is carried over to the visualizations. We found that when comparing visualizations of words that share more phones against words that share fewer (or no) phones, users always (100\%) found shared phones to be more similar for the CelebA visualizations while they were more similar 95\% of the time for the MNIST visualizations. In addition, when trying to assess the similarity between phone pairs that do or do not share a common phone, users reported phone pairs that do share a phone as more similar 73.3\% and 69.4\% of the time for the MNIST and CelebA visualizations, respectively.
}

\subsection{Ablation Study}
\label{sec:throughput}
\paragraph{Information Throughput.}

\begin{table}[]
\resizebox{0.98\linewidth}{!}{
\centering
\begin{tabular}{llr}
\toprule
Audio models                                               & Visual models                       &  SNR(dB)                    \\
\midrule
\multirow{1}{*}{Audio PCA}                              & Visual PCA                           & 23.37                    \\
\midrule
\multirow{4}{*}{SpeechVAE \cite{hsu2017learning} } 
                                             & DFC-VAE on CelebA  & 1.65   \\
                                             & DFC-VAE on MNIST   & 2.01    \\ 
                                             & DFC-VAE on CelebA (refined w/ $\mathcal{L}_{cycle}$)  & \textbf{4.43}   \\
                                             & DFC-VAE on MNIST (refiend w/ $\mathcal{L}_{cycle}$)   & 0.78  \\
                                           \cline{1-3}
\multirow{5}{*}{\begin{tabular}[c]{@{}l@{}}SpeechVAE w/ $\mathcal{L}_{p, \log},$ \\ $\mathcal{L}_{rr}$, dim=256  \end{tabular}} 

                                       & DFC-VAE on CelebA & 0.84    \\
                                       & DFC-VAE on MNIST  & 0.81   \\
                                     & DFC-VAE on CelebA (refined w/ $\mathcal{L}_{cycle}$)  & \textbf{4.16}   \\
                                     & DFC-VAE on MNIST (refiend w/ $\mathcal{L}_{cycle}$)   & 3.68  \\
                                     &  Soft-Intro VAE on CelebA\_HR (refined w/ $\mathcal{L}_{cycle}$)                           & 2.01  \\

\bottomrule
\end{tabular}}
\caption{\textbf{Throughput Analysis.} The throughput estimated from audio to video %
shows that the cycle consistency improves greatly while additional constraints $\mathcal{L}_{p, \log}$, and $\mathcal{L}_{rr}$ incur only a small dip for improving smoothness. \cj{$dim$ indicates the dimension of the audio latent space.} Results are consistent over different models. \looseness=-1}
\label{tab:audio_and_video}
\end{table}
\comment{
\begin{figure}[t]
\floatbox[{\capbeside\thisfloatsetup{capbesideposition={right,top},capbesidewidth=0.45\textwidth}}]{figure}[\FBwidth]
{\caption{\textbf{Throughput.} A visualization of the lower bound on the information throughput when mapping from audio to video and back.
}\label{fig:throughput}}
{\includegraphics[width=0.5\textwidth, trim=0 11.5cm 5cm 0,clip]{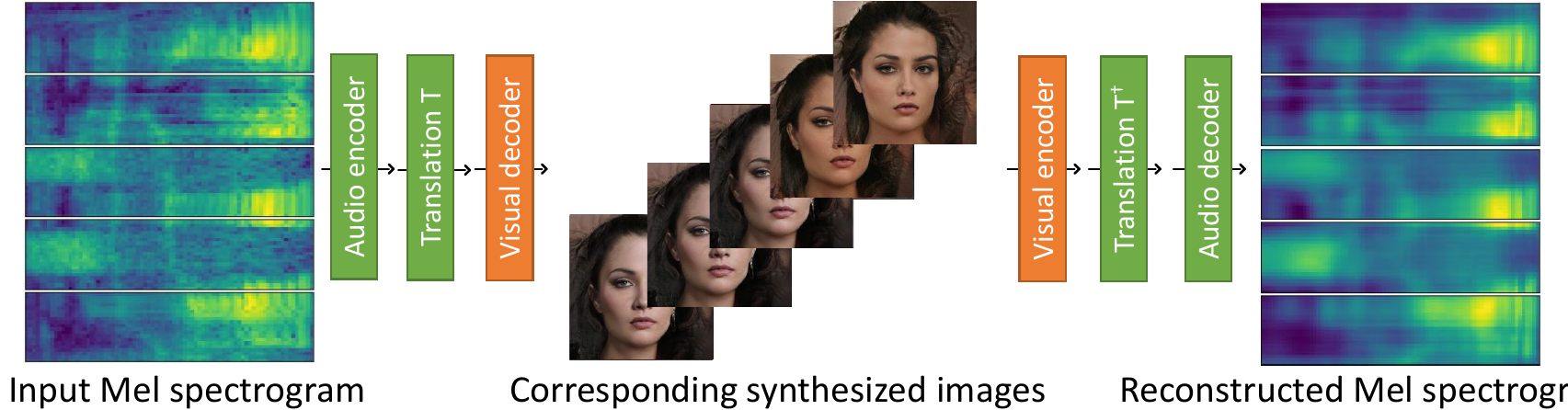}}

\end{figure}
}
 
\begin{figure}[t!]
	\centering
	{\includegraphics[width=0.98\textwidth, trim=0 11.5cm 3cm 0,clip]{images/throughput.pdf}}
	\caption{\textbf{Throughput.} A visualization of the lower bound on the information throughput when mapping from audio to video and back. \looseness=-1}
	\label{fig:throughput}
\end{figure}

It is difficult to quantify the information throughput from audio to
video as no ground truth is available in our setting. We propose to use the learned encoder and decoder to map from audio to video and back.
The distance of the starting point to the reconstructed audio gives a lower bound on the loss of information, akin to CycleGAN.
Fig.~\ref{fig:throughput} gives an example and Table~\ref{tab:audio_and_video} summarizes relations quantitatively. 
The difference can also be analyzed qualitatively by listening and comparing the original and reconstructed audio samples, which are still recognizable for our full model. 
We found this throughput an important measure that correlated with performance in the human studies and enabled us to tune hyperparameters without performing an expensive study for each configuration.
Note that the information throughput rivals other constraints such as smoothness. The goal is therefore to strike the best compromise.
For instance, PCA attains the highest reconstruction accuracy but has poor smoothness (Table~\ref{tab:audio_only}) and does not strike along the other dimensions.
The disentangled space (lower half of Table~\ref{tab:audio_and_video}) has a relatively low effect on the SNR, a reduction from $4.43$ to $4.16$, while providing improved interpretability. MNIST proved less stable to train and does not fair well with the cycle loss, perhaps due to a lower dimensionality that mismatches with the higher-dimensional audio encoding.
The high-res CelebA model yields a lower SNR, likely because information is lost in the deeper network and the smoothness loss affects high-frequency details differently.\looseness=-1

\comment{
We further analyze the effect of the additional constraints on the audio autoencoder by measuring its throughput individually.
Out of the models with smoothness, $\mathcal{L}_{p, \log}$ with $\lambda_p=10^3$ attains the highest reconstruction quality (SNR) and lowest latent space velocity calculated with finite differences. The disentanglement reduces audio reconstruction significantly. However, this has limited effect on the overall throughput, as analyzed before, likely due to bottlenecks in the visual model dominating the overall accuracy.}

\parag{Temporal Smoothness Effectiveness.}
Mapping from audio to video with VAEs without constraints and PCA leads to choppy results. Table~\ref{tab:audio_only} shows that this corresponds to mean latent space velocities above 300 s$^{-1}$. It shows that the tested simpler solutions are insufficient. We visualize the gain in smoothness in Fig.~\ref{fig:embedding_curves} by plotting the high-dimensional latent trajectory embedded into three dimensions using 
multi-dimensional scaling (MDS) \cite{cox2008multidimensional}. The gain is similar for all of the proposed variants. The supplemental videos show how the smoothness eases information perception.

\parag{Visual Domain Influence.}
Earlier work~\cite{chernoff1973use,jacob1976face} suggests that faces are suitable for representing high-dimensional data. Our experiments support this finding for the task at hand. 
Faces performed better than digits in our human study as facial models are richer in information. This is confirmed quantitatively by our throughput experiments, the SNR of the jointly trained CelebA models is larger than the MNIST models (visualized in Fig.~\ref{fig:phonemes content}). As expected, the human study also reveals that high-resolution faces work better than low-resolution faces. \looseness=-1
\comment{
\begin{figure}[t]
\floatbox[{\capbeside\thisfloatsetup{capbesideposition={right,top},capbesidewidth=0.45\textwidth}}]{figure}[\FBwidth]
{\caption{\textbf{Smoothness evaluation} by plotting the latent embedding of an audio snipped in 3D via dimensionality reduction. For PCA and w/o smoothness constraint, consecutive audio embeddings are scattered.}\label{fig:embedding_curves}}
{\includegraphics[width=0.5\textwidth, trim=0 0.2cm 0 0,clip]{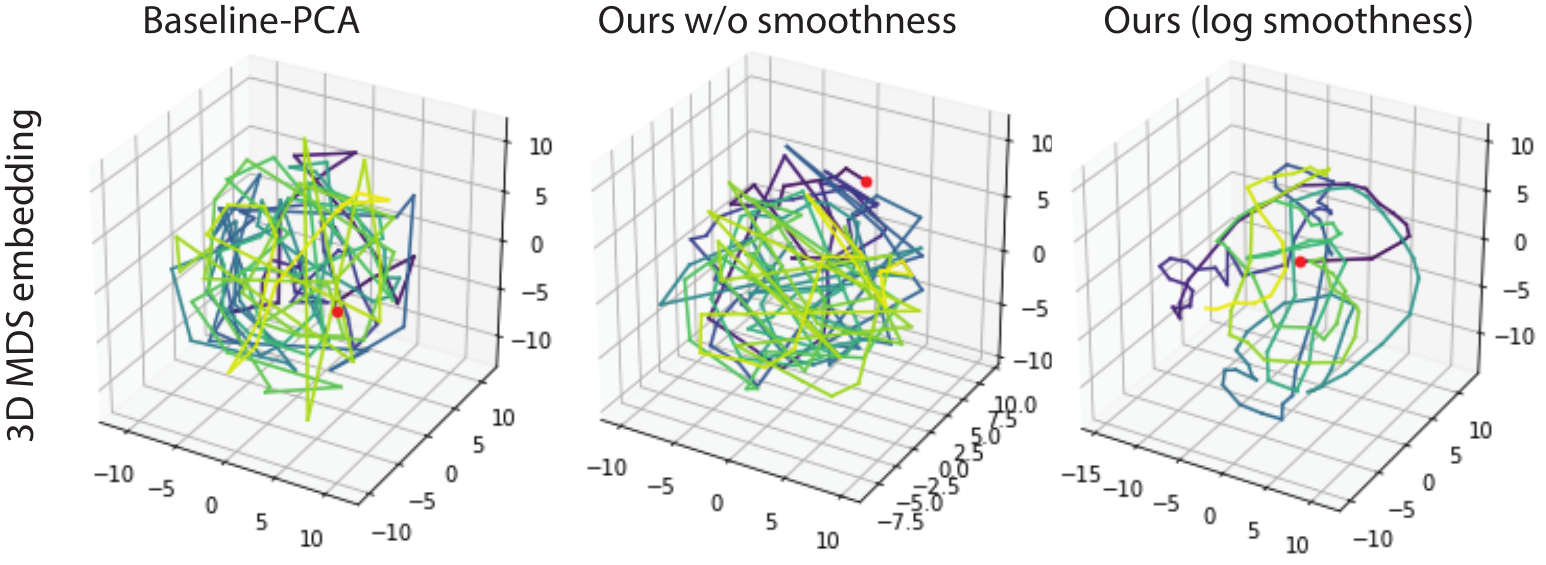}}

\end{figure}
}

\begin{figure}[t]
	\centering
	\includegraphics[width=0.95\linewidth,trim=0 0.2cm 0 0,clip]{images/smoothness_MDS4.pdf}
	\caption{\textbf{Smoothness evaluation} by plotting the latent embedding of an audio snipped in 3D via dimensionality reduction. For PCA and w/o smoothness constraint, consecutive audio embeddings are scattered. \looseness=-1}
	\label{fig:embedding_curves}
\end{figure}

\begin{table}[]
\resizebox{0.98\linewidth}{!}{
\centering
\begin{tabular}{lrrr}
\toprule
Audio models   &   SNR (dB) & Velocity ($\mathrm{s}^{-1}$)  &  Acc.($\mathrm{s}^{-2}$)   \\ 
\midrule
Audio PCA  &   \textbf{23.37}   &   329.02    &   13395.11  \\
SpeechVAE\cite{hsu2017learning}  &   21.89    &   \textbf{280.09}    &   \textbf{11648.17} \\
\cline{1-4}
{SpeechVAE w/ $\mathcal{L}_{p, \log}$ } & \textbf{19.89} & 108.89 & 3052.19 \\
{SpeechVAE w/ $\mathcal{L}_{rr}$ } & 7.73 & 80.80 & 2947.70 \\
{SpeechVAE w/ $\mathcal{L}_{rr}$,$\mathcal{L}_{p, \log}$ } 
                    & 6.20 & \textbf{59.78} & \textbf{1530.81} \\
{SpeechVAE w/$\mathcal{L}_{rr}$, $\mathcal{L}_{p, \log}$, dim=256 }
                    & 6.28 & 65.63 & 1757.95 \\
\bottomrule\\
$ $\\
$ $\\
\end{tabular}}%
\caption{We quantify the improvement in velocity and acceleration in relation to audio encode throughput in SNR on the best performing model, DFC-VAE on CelebA.}
\label{tab:audio_only}
\vskip -0.4in
\end{table}

\section{Limitations and Future Work}

When presented with our visualizations the first time, users are unsure of what features they should focus on to distinguish. Our Study II shows that this improves with learning but similar sounds, such as \emph{that} and \emph{there} remain somewhat difficult to distinguish. Which is no surprise as the input audio is similar, but there is room to improve the expressiveness and make the visual language more intuitive. Moreover, the visualization of environment sounds lose higher frequency components since the audio models are trained on a speech dataset.
To create a more natural mapping, we plan to map to lip motion for spoken sounds that are captured in the lip motion while maintaining the proposed general encoding for those that are not and for environment sounds.

\comment{
* no semantics.
* no temporal smoothness since trained on images not videos.
* not validated for deaf people.

Not just totally deaf persons, also many dyslexic deficits can be associated to missing subtleties in hearing. Perhaps, visual representations could help. Further life science studies are needed to explore these possible directions enabled by our \ours{} translation system.

Future work ideas:
• Language specific models or general ones? (different for mandarin and English?)
* real-time?
* supervised with phenom labels
* higher capacity / modern video generators
* stereo video output / 3D output
* stereo audio input
* low latency generation
* different temporal granularity
* using more sophisticated perceptual metrics
• Medical / psychological / linguistic studies
• How can models be updated after deployment. A user learns the meaning of images,
replacement of the image synthesis model would require learning the correspondence anew
with each update.

* disentanglement of style and content
* joint latent space with SWD of different batch size
* visualize as hand written use letters/Phonetic Alphabet 
    - I feel like figures are easier to

* Additional users studies can feature live demos, and having participants trying to decode sequences of video via pronunciation feedback
* perhaps adding some constraints /loss on the visuals might be good i.e. that different sounds should look different

}

\section{Conclusion}

We presented \ours{}, a learning-based approach for visualizing audio, including speech and sounds, via generative models trained in the absence of paired examples.
It can visualize sounds on faces and digits and outperforms all baselines in distinctiveness, including state-of-the-art lip synchronization and spectrograms as used in current assistive systems. Our face visualizations retain more information than lip motion and are easier to parse than spectrograms.
Our proof of concept \ours{} demo shows the feasibility of visualizing speech in real-time. We hope that our sensory substitution approach will catalyze the development of future tools for assisting people with auditorial handicaps. %

\comment{
\section*{Broader Impact}
When used as an assistive tool, \ours{} has the potential to have a positive impact on society by assisting and better connecting people with hearing impairment. However, such technology should be treated with care. It will require careful clinical trials to rule out negative effects; as with any tool or mechanism used for medical training, particularly when working with children. We train and test our model on spoken English and general sounds, but biases in the data distribution of such datasets can carry over to our trained model. For example, the former dataset only represents American voices, where male speakers are over-represented. One has to be careful to ensure that such biases are mitigated before such a tool can be reliably used by the public.
However, since our method is self-supervised, only speech samples are required as input and can be re-trained for languages for which no annotated training sets exist.

\begin{ack}
We thank all participants of the user study for their time and contribution and Farnoosh Javadi for improving this manuscript.
\end{ack}
}

\section{Acknowledgement}
We thank Ipek Oruk for her valuable comments and all participants of the user study for their time and contribution.
This work was supported by the Compute Canada GPU servers.

{\small
\bibliographystyle{ieee_fullname}
\bibliography{tex/references}

\begin{thebibliography}{10}\itemsep=-1pt

\bibitem{aguirre1998area}
Geoffrey~K Aguirre, E Zarahn, and M D’esposito.
\newblock An area within human ventral cortex sensitive to “building”
  stimuli: evidence and implications.
\newblock {\em Neuron}, 21(2):373--383, 1998.

\bibitem{briot2017deep}
Jean-Pierre Briot, Ga{\"e}tan Hadjeres, and Fran{\c{c}}ois-David Pachet.
\newblock Deep learning techniques for music generation--a survey.
\newblock {\em arXiv preprint arXiv:1709.01620}, 2017.

\bibitem{cappelletta2012phoneme}
Luca Cappelletta and Naomi Harte.
\newblock Phoneme-to-viseme mapping for visual speech recognition.
\newblock In {\em ICPRAM (2)}, pages 322--329, 2012.

\bibitem{chen2018lip}
Lele Chen, Zhiheng Li, Ross~K Maddox, Zhiyao Duan, and Chenliang Xu.
\newblock Lip movements generation at a glance.
\newblock In {\em Proceedings of the European Conference on Computer Vision
  (ECCV)}, pages 520--535, 2018.

\bibitem{chen2017deepcrossmodal}
Lele Chen, Sudhanshu Srivastava, Zhiyao Duan, and Chenliang Xu.
\newblock Deep cross-modal audio-visual generation.
\newblock In {\em Proceedings of the on Thematic Workshops of ACM Multimedia
  2017}, Thematic Workshops '17, page 349–357, New York, NY, USA, 2017.
  Association for Computing Machinery.

\bibitem{chen2019soundtovisual}
Lele Chen, Haitian Zheng, Ross Maddox, Zhiyao Duan, and Chenliang Xu.
\newblock Sound to visual: Hierarchical cross-modal talking face generation.
\newblock In {\em Proceedings of the IEEE/CVF Conference on Computer Vision and
  Pattern Recognition (CVPR) Workshops}, June 2019.

\bibitem{chernoff1973use}
Herman Chernoff.
\newblock The use of faces to represent points in k-dimensional space
  graphically.
\newblock {\em Journal of the American statistical Association},
  68(342):361--368, 1973.

\bibitem{christensen2020batvision}
Jesper~Haahr Christensen, Sascha Hornauer, and X~Yu Stella.
\newblock Batvision: Learning to see 3d spatial layout with two ears.
\newblock In {\em 2020 IEEE International Conference on Robotics and Automation
  (ICRA)}, pages 1581--1587. IEEE, 2020.

\bibitem{chu2017cyclegan}
Casey Chu, Andrey Zhmoginov, and Mark Sandler.
\newblock Cyclegan, a master of steganography.
\newblock {\em arXiv preprint arXiv:1712.02950}, 2017.

\bibitem{chung2017you}
Joon~Son Chung, Amir Jamaludin, and Andrew Zisserman.
\newblock You said that?
\newblock {\em arXiv preprint arXiv:1705.02966}, 2017.

\bibitem{cox2008multidimensional}
Michael~AA Cox and Trevor~F Cox.
\newblock Multidimensional scaling.
\newblock In {\em Handbook of data visualization}, pages 315--347. Springer,
  2008.

\bibitem{daniel2021soft}
Tal Daniel and Aviv Tamar.
\newblock Soft-introvae: Analyzing and improving the introspective variational
  autoencoder.
\newblock In {\em Proceedings of the IEEE/CVF Conference on Computer Vision and
  Pattern Recognition}, pages 4391--4400, 2021.

\bibitem{dong2018musegan}
Hao-Wen Dong, Wen-Yi Hsiao, Li-Chia Yang, and Yi-Hsuan Yang.
\newblock Musegan: Multi-track sequential generative adversarial networks for
  symbolic music generation and accompaniment.
\newblock In {\em Thirty-Second AAAI Conference on Artificial Intelligence},
  2018.

\bibitem{duarte2019wav2pix}
Amanda Duarte, Francisco Roldan, Miquel Tubau, Janna Escur, Santiago Pascual,
  Amaia Salvador, Eva Mohedano, Kevin McGuinness, Jordi Torres, and Xavier
  Giro-i Nieto.
\newblock Wav2pix: speech-conditioned face generation using generative
  adversarial networks.
\newblock In {\em IEEE International Conference on Acoustics, Speech and Signal
  Processing (ICASSP)}, volume~3, 2019.

\bibitem{eberhardt1993omar}
Silvio~P Eberhardt, Lynne~E Bernstein, David~C Coulter, and Laura~A Hunckler.
\newblock Omar a haptic display for speech perception by deaf and deaf-blind
  individuals.
\newblock In {\em Proceedings of IEEE Virtual Reality Annual International
  Symposium}, pages 195--201. IEEE, 1993.

\bibitem{elssmann1987speech}
Sharon~F Elssmann and Jean~E Maki.
\newblock Speech spectrographic display: use of visual feedback by
  hearing-impaired adults during independent articulation practice.
\newblock {\em American Annals of the Deaf}, pages 276--279, 1987.

\bibitem{fernandez2017optimizing}
Adriana Fernandez-Lopez and Federico~M Sukno.
\newblock Optimizing phoneme-to-viseme mapping for continuous lip-reading in
  spanish.
\newblock In {\em International Joint Conference on Computer Vision, Imaging
  and Computer Graphics}, pages 305--328. Springer, 2017.

\bibitem{garofolo1993timit}
John~S Garofolo.
\newblock Timit acoustic phonetic continuous speech corpus.
\newblock {\em Linguistic Data Consortium, 1993}, 1993.

\bibitem{goldish1974optacon}
Louis~H Goldish and Harry~E Taylor.
\newblock The optacon: A valuable device for blind persons.
\newblock {\em Journal of Visual Impairment \& Blindness}, 68(2):49--56, 1974.

\bibitem{goodfellow2014generative}
Ian Goodfellow, Jean Pouget-Abadie, Mehdi Mirza, Bing Xu, David Warde-Farley,
  Sherjil Ozair, Aaron Courville, and Yoshua Bengio.
\newblock Generative adversarial nets.
\newblock In {\em Advances in neural information processing systems}, pages
  2672--2680, 2014.

\bibitem{hao2018cmcgan}
Wangli Hao, Zhaoxiang Zhang, and He Guan.
\newblock Cmcgan: A uniform framework for cross-modal visual-audio mutual
  generation.
\newblock In {\em Thirty-Second AAAI Conference on Artificial Intelligence},
  2018.

\bibitem{hardcastle1991visual}
William~J Hardcastle, Fiona~E Gibbon, and Wilf Jones.
\newblock Visual display of tongue-palate contact: electropalatography in the
  assessment and remediation of speech disorders.
\newblock {\em International Journal of Language \& Communication Disorders},
  26(1):41--74, 1991.

\bibitem{hiasa2018cross}
Yuta Hiasa, Yoshito Otake, Masaki Takao, Takumi Matsuoka, Kazuma Takashima,
  Aaron Carass, Jerry~L Prince, Nobuhiko Sugano, and Yoshinobu Sato.
\newblock Cross-modality image synthesis from unpaired data using cyclegan.
\newblock In {\em International workshop on simulation and synthesis in medical
  imaging}, pages 31--41. Springer, 2018.

\bibitem{Hou2017}
X. {Hou}, L. {Shen}, K. {Sun}, and G. {Qiu}.
\newblock Deep feature consistent variational autoencoder.
\newblock In {\em 2017 IEEE Winter Conference on Applications of Computer
  Vision (WACV)}, pages 1133--1141, 2017.

\bibitem{hsu2017learning}
Wei-Ning Hsu, Yu Zhang, and James Glass.
\newblock Learning latent representations for speech generation and
  transformation.
\newblock In {\em Interspeech}, pages 1273--1277, 2017.

\bibitem{Hu_2019_CVPR}
Di Hu, Dong Wang, Xuelong Li, Feiping Nie, and Qi Wang.
\newblock Listen to the image.
\newblock In {\em Proceedings of the IEEE/CVF Conference on Computer Vision and
  Pattern Recognition (CVPR)}, June 2019.

\bibitem{huang2017arbitrary}
Xun Huang and Serge Belongie.
\newblock Arbitrary style transfer in real-time with adaptive instance
  normalization.
\newblock In {\em Proceedings of the IEEE International Conference on Computer
  Vision}, pages 1501--1510, 2017.

\bibitem{ibm2004speech}
IBM.
\newblock Speech viewer iii, 2004.

\bibitem{jacob1976face}
Robert~JK Jacob and Howard~E Egeth.
\newblock The face as a data display.
\newblock {\em Human Factors}, 18(2):189--200, 1976.

\bibitem{jamaludin2019you}
Amir Jamaludin, Joon~Son Chung, and Andrew Zisserman.
\newblock You said that?: Synthesising talking faces from audio.
\newblock {\em International Journal of Computer Vision}, 127(11):1767--1779,
  2019.

\bibitem{jha2018disentangling}
Ananya~Harsh Jha, Saket Anand, Maneesh Singh, and VSR Veeravasarapu.
\newblock Disentangling factors of variation with cycle-consistent variational
  auto-encoders.
\newblock In {\em European Conference on Computer Vision}, pages 829--845.
  Springer, 2018.

\bibitem{kamper2019truly}
Herman Kamper.
\newblock Truly unsupervised acoustic word embeddings using weak top-down
  constraints in encoder-decoder models.
\newblock In {\em ICASSP 2019-2019 IEEE International Conference on Acoustics,
  Speech and Signal Processing (ICASSP)}, pages 6535--3539. IEEE, 2019.

\bibitem{kanwisher1999fusiform}
Nancy Kanwisher, Damian Stanley, and Alison Harris.
\newblock The fusiform face area is selective for faces not animals.
\newblock {\em Neuroreport}, 10(1):183--187, 1999.

\bibitem{karras2017audio}
Tero Karras, Timo Aila, Samuli Laine, Antti Herva, and Jaakko Lehtinen.
\newblock Audio-driven facial animation by joint end-to-end learning of pose
  and emotion.
\newblock {\em ACM Transactions on Graphics (TOG)}, 36(4):1--12, 2017.

\bibitem{karras2017progressive}
Tero Karras, Timo Aila, Samuli Laine, and Jaakko Lehtinen.
\newblock Progressive growing of gans for improved quality, stability, and
  variation.
\newblock {\em arXiv preprint arXiv:1710.10196}, 2017.

\bibitem{karras2019style}
Tero Karras, Samuli Laine, and Timo Aila.
\newblock A style-based generator architecture for generative adversarial
  networks.
\newblock In {\em Proceedings of the IEEE Conference on Computer Vision and
  Pattern Recognition}, pages 4401--4410, 2019.

\bibitem{katz2015visual}
William~F Katz and Sonya Mehta.
\newblock Visual feedback of tongue movement for novel speech sound learning.
\newblock {\em Frontiers in human neuroscience}, 9:612, 2015.

\bibitem{kingma2013auto}
Diederik~P Kingma and Max Welling.
\newblock Auto-encoding variational bayes.
\newblock {\em arXiv preprint arXiv:1312.6114}, 2013.

\bibitem{kr2019towards}
Prajwal KR, Rudrabha Mukhopadhyay, Jerin Philip, Abhishek Jha, Vinay
  Namboodiri, and CV Jawahar.
\newblock Towards automatic face-to-face translation.
\newblock In {\em Proceedings of the 27th ACM International Conference on
  Multimedia}, pages 1428--1436, 2019.

\bibitem{kroger2010audiovisual}
Bernd~J Kr{\"o}ger, Peter Birkholz, R{\"u}diger Hoffmann, and Helen Meng.
\newblock Audiovisual tools for phonetic and articulatory visualization in
  computer-aided pronunciation training.
\newblock In {\em Development of multimodal interfaces: Active listening and
  synchrony}, pages 337--345. Springer, 2010.

\bibitem{lecun1998gradient}
Yann LeCun, L{\'e}on Bottou, Yoshua Bengio, and Patrick Haffner.
\newblock Gradient-based learning applied to document recognition.
\newblock {\em Proceedings of the IEEE}, 86(11):2278--2324, 1998.

\bibitem{levis2004teaching}
John Levis and Lucy Pickering.
\newblock Teaching intonation in discourse using speech visualization
  technology.
\newblock {\em System}, 32(4):505--524, 2004.

\bibitem{lidestam2006visual}
Bj{\"o}rn Lidestam and Jonas Beskow.
\newblock Visual phonemic ambiguity and speechreading.
\newblock {\em Journal of Speech, Language, and Hearing Research}, 2006.

\bibitem{liu2015faceattributes}
Ziwei Liu, Ping Luo, Xiaogang Wang, and Xiaoou Tang.
\newblock Deep learning face attributes in the wild.
\newblock In {\em Proceedings of International Conference on Computer Vision
  (ICCV)}, December 2015.

\bibitem{maidenbaum2014sensory}
Shachar Maidenbaum, Sami Abboud, and Amir Amedi.
\newblock Sensory substitution: Closing the gap between basic research and
  widespread practical visual rehabilitation.
\newblock {\em Neuroscience \& Biobehavioral Reviews}, 41:3--15, 2014.

\bibitem{mama2021nwt}
Rayhane Mama, Marc~S. Tyndel, Hashiam Kadhim, Cole Clifford, and Ragavan
  Thurairatnam.
\newblock Nwt: Towards natural audio-to-video generation with representation
  learning, 2021.

\bibitem{musk2019integrated}
Elon Musk et~al.
\newblock An integrated brain-machine interface platform with thousands of
  channels.
\newblock {\em Journal of medical Internet research}, 21(10):e16194, 2019.

\bibitem{newman2010limitations}
Jacob~L Newman, Barry-John Theobald, and Stephen~J Cox.
\newblock Limitations of visual speech recognition.
\newblock In {\em Auditory-Visual Speech Processing 2010}, 2010.

\bibitem{noda2015audio}
Kuniaki Noda, Yuki Yamaguchi, Kazuhiro Nakadai, Hiroshi~G Okuno, and Tetsuya
  Ogata.
\newblock Audio-visual speech recognition using deep learning.
\newblock {\em Applied Intelligence}, 42(4):722--737, 2015.

\bibitem{oster1995teaching}
AM {\"O}ster.
\newblock Teaching speech skills to deaf children by computer-based speech
  training.
\newblock {\em STL-Quarterly Progress and Status Report}, 36(4):67--75, 1995.

\bibitem{park1994integrated}
Sang~H Park, Dong~J Kim, Jae~H Lee, and Tae~S Yoon.
\newblock Integrated speech training system for hearing impaired.
\newblock {\em IEEE Transactions on Rehabilitation Engineering}, 2(4):189--196,
  1994.

\bibitem{piczak2015esc}
Karol~J Piczak.
\newblock Esc: Dataset for environmental sound classification.
\newblock In {\em Proceedings of the 23rd ACM international conference on
  Multimedia}, pages 1015--1018, 2015.

\bibitem{povey2011kaldi}
Daniel Povey, Arnab Ghoshal, Gilles Boulianne, Lukas Burget, Ondrej Glembek,
  Nagendra Goel, Mirko Hannemann, Petr Motlicek, Yanmin Qian, Petr Schwarz,
  et~al.
\newblock The kaldi speech recognition toolkit.
\newblock In {\em IEEE 2011 workshop on automatic speech recognition and
  understanding}, number CONF. IEEE Signal Processing Society, 2011.

\bibitem{prajwal2020Lip}
K~R Prajwal, Rudrabha Mukhopadhyay, Vinay~P. Namboodiri, and C.V. Jawahar.
\newblock A lip sync expert is all you need for speech to lip generation in the
  wild.
\newblock In {\em Proceedings of the 28th ACM International Conference on
  Multimedia}, MM '20, page 484–492, New York, NY, USA, 2020. Association for
  Computing Machinery.

\bibitem{qian2019autovc}
Kaizhi Qian, Yang Zhang, Shiyu Chang, Xuesong Yang, and Mark Hasegawa-Johnson.
\newblock Autovc: Zero-shot voice style transfer with only autoencoder loss.
\newblock In {\em International Conference on Machine Learning}, pages
  5210--5219. PMLR, 2019.

\bibitem{qiu2018image}
Yue Qiu and Hirokatsu Kataoka.
\newblock Image generation associated with music data.
\newblock In {\em Proceedings of the IEEE Conference on Computer Vision and
  Pattern Recognition Workshops}, pages 2510--2513, 2018.

\bibitem{sadoughi2019speech}
Najmeh Sadoughi and Carlos Busso.
\newblock Speech-driven expressive talking lips with conditional sequential
  generative adversarial networks.
\newblock {\em IEEE Transactions on Affective Computing}, 2019.

\bibitem{shlizerman2018audio}
Eli Shlizerman, Lucio Dery, Hayden Schoen, and Ira Kemelmacher-Shlizerman.
\newblock Audio to body dynamics.
\newblock In {\em Proceedings of the IEEE Conference on Computer Vision and
  Pattern Recognition}, pages 7574--7583, 2018.

\bibitem{sperling1960information}
George Sperling.
\newblock The information available in brief visual presentations.
\newblock {\em Psychological monographs: General and applied}, 74(11):1, 1960.

\bibitem{stewart1976a}
Larkin~W Stewart~L and Houde R.
\newblock A real-time sound spectrograph with implications for speech training
  for the deaf.
\newblock In {\em IEEE International Conference Accoustics Speech and Signal
  Processing}, 1976.

\bibitem{suwajanakorn2017synthesizing}
Supasorn Suwajanakorn, Steven~M Seitz, and Ira Kemelmacher-Shlizerman.
\newblock Synthesizing obama: learning lip sync from audio.
\newblock {\em ACM Transactions on Graphics (TOG)}, 36(4):1--13, 2017.

\bibitem{taigman2016unsupervised}
Yaniv Taigman, Adam Polyak, and Lior Wolf.
\newblock Unsupervised cross-domain image generation.
\newblock {\em arXiv preprint arXiv:1611.02200}, 2016.

\bibitem{takahashi2018instrudive}
Takumi Takahashi, Satoru Fukayama, and Masataka Goto.
\newblock Instrudive: A music visualization system based on automatically
  recognized instrumentation.
\newblock In {\em ISMIR}, pages 561--568, 2018.

\bibitem{tarr2000ffa}
Michael~J Tarr and Isabel Gauthier.
\newblock Ffa: a flexible fusiform area for subordinate-level visual processing
  automatized by expertise.
\newblock {\em Nature neuroscience}, 3(8):764--769, 2000.

\bibitem{taylor2017deep}
Sarah Taylor, Taehwan Kim, Yisong Yue, Moshe Mahler, James Krahe,
  Anastasio~Garcia Rodriguez, Jessica Hodgins, and Iain Matthews.
\newblock A deep learning approach for generalized speech animation.
\newblock {\em ACM Transactions on Graphics (TOG)}, 36(4):1--11, 2017.

\bibitem{tian2019latent}
Yingtao Tian and Jesse Engel.
\newblock Latent translation: Crossing modalities by bridging generative
  models.
\newblock {\em arXiv preprint arXiv:1902.08261}, 2019.

\bibitem{tmenova2019cyclegan}
Oleksandra Tmenova, R{\'e}mi Martin, and Luc Duong.
\newblock Cyclegan for style transfer in x-ray angiography.
\newblock {\em International journal of computer assisted radiology and
  surgery}, 14(10):1785--1794, 2019.

\bibitem{vougioukas2018end}
Konstantinos Vougioukas, Stavros Petridis, and Maja Pantic.
\newblock End-to-end speech-driven facial animation with temporal gans.
\newblock {\em arXiv preprint arXiv:1805.09313}, 2018.

\bibitem{wan2019towards}
Chia-Hung Wan, Shun-Po Chuang, and Hung-Yi Lee.
\newblock Towards audio to scene image synthesis using generative adversarial
  network.
\newblock In {\em ICASSP 2019-2019 IEEE International Conference on Acoustics,
  Speech and Signal Processing (ICASSP)}, pages 496--500. IEEE, 2019.

\bibitem{wiles2018x2face}
Olivia Wiles, A Sophia~Koepke, and Andrew Zisserman.
\newblock X2face: A network for controlling face generation using images,
  audio, and pose codes.
\newblock In {\em Proceedings of the European Conference on Computer Vision
  (ECCV)}, pages 670--686, 2018.

\bibitem{xu2008speech}
Wang Xu, Xue Lifang, Yang Dan, and Han Zhiyan.
\newblock Speech visualization based on robust self-organizing map (rsom) for
  the hearing impaired.
\newblock In {\em 2008 International Conference on BioMedical Engineering and
  Informatics}, volume~2, pages 506--509. IEEE, 2008.

\bibitem{yang2010speech}
Dan Yang, Bin Xu, and Xu Wang.
\newblock Speech visualization based on improved spectrum for deaf children.
\newblock In {\em 2010 Chinese Control and Decision Conference}, pages
  4377--4380. IEEE, 2010.

\bibitem{yook2020many}
Dongsuk Yook, Seong-Gyun Leem, Keonnyeong Lee, and In-Chul Yoo.
\newblock Many-to-many voice conversion using cycle-consistent variational
  autoencoder with multiple decoders.
\newblock In {\em Proc. Odyssey 2020 The Speaker and Language Recognition
  Workshop}, pages 215--221, 2020.

\bibitem{yoshii2008music}
Kazuyoshi Yoshii and Masataka Goto.
\newblock Music thumbnailer: Visualizing musical pieces in thumbnail images
  based on acoustic features.
\newblock In {\em ISMIR}, pages 211--216, 2008.

\bibitem{yuan2019speechlens}
Linping Yuan, Yuanzhe Chen, Siwei Fu, Aoyu Wu, and Huamin Qu.
\newblock Speechlens: A visual analytics approach for exploring speech
  strategies with textural and acoustic features.
\newblock In {\em 2019 IEEE International Conference on Big Data and Smart
  Computing (BigComp)}, pages 1--8. IEEE, 2019.

\bibitem{zaccagnini1993effects}
Cindy~M Zaccagnini and Shirin~D Antia.
\newblock Effects of multisensory speech training and visual phonics on speech
  production of a hearing-impaired child.
\newblock {\em Journal of Childhool Communication Disorders}, 15(2):3--8, 1993.

\bibitem{zhou2019talking}
Hang Zhou, Yu Liu, Ziwei Liu, Ping Luo, and Xiaogang Wang.
\newblock Talking face generation by adversarially disentangled audio-visual
  representation.
\newblock In {\em Proceedings of the AAAI Conference on Artificial
  Intelligence}, volume~33, pages 9299--9306, 2019.

\bibitem{zhou2020makeittalk}
Yang Zhou, Xintong Han, Eli Shechtman, Jose Echevarria, Evangelos Kalogerakis,
  and Dingzeyu Li.
\newblock Makelttalk: Speaker-aware talking-head animation.
\newblock {\em ACM Trans. Graph.}, 39(6), Nov. 2020.

\bibitem{zhu2021deep}
Hao Zhu, Man-Di Luo, Rui Wang, Ai-Hua Zheng, and Ran He.
\newblock Deep audio-visual learning: A survey.
\newblock {\em International Journal of Automation and Computing},
  18(3):351--376, 2021.

\bibitem{zhu2017unpaired}
Jun-Yan Zhu, Taesung Park, Phillip Isola, and Alexei~A Efros.
\newblock Unpaired image-to-image translation using cycle-consistent
  adversarial networks.
\newblock In {\em Proceedings of the IEEE international conference on computer
  vision}, pages 2223--2232, 2017.

\end{thebibliography}
}

\end{document}


\title{Supplemental Document\\AudioViewer: Learning to Visualize Sounds}

\maketitle
This document supplements the main paper by providing more analysis with the related works, more implementation details, content and style disentanglement, an additional ablation study regarding the translation network $T$, an additional analysis on the human study, and training details.  We also prepared an additional supplemental video that contains audio and video snippets that can not be played in a conventional PDF. These give additional qualitative results for spoken sentences and environment sounds.

\textbf{Licenses.} The used audio datesets TIMIT dataset~\cite{garofolo1993timit} and ESC-50 dataset~\cite{piczak2015esc} in our experiments are public. TIMIT dataset~\cite{garofolo1993timit} is under the terms of LDC User Agreement for Non-Members license and ESC-50 dataset~\cite{piczak2015esc} is under the terms of the Creative Commons Attribution Non-Commercial license.

\textbf{Risk mitigation and scope.} 
Sensory substitution bears non-negligible risks. Our comparative study design is approved by our IRB to have a low risk.
Whether an entire language can be learned will require psychophysical studies controlled by domain experts to mitigate the risk of side effects on long-term participants.

\comment{
\subsection{Style}

Papers to be submitted to NeurIPS 2020 must be prepared according to the
instructions presented here. Papers may only be up to eight pages long,
including figures. Additional pages \emph{containing only a section on the broader impact, acknowledgments and/or cited references} are allowed. Papers that exceed eight pages of content will not be reviewed, or in any other way considered for
presentation at the conference.

\paragraph{Preprint option}
If you wish to post a preprint of your work online, e.g., on arXiv, using the
NeurIPS style, please use the \verb+preprint+ option. This will create a
nonanonymized version of your work with the text ``Preprint. Work in progress.''
in the footer. This version may be distributed as you see fit. Please \textbf{do
  not} use the \verb+final+ option, which should \textbf{only} be used for
papers accepted to NeurIPS.

At submission time, please omit the \verb+final+ and \verb+preprint+
options. This will anonymize your submission and add line numbers to aid
review. Please do \emph{not} refer to these line numbers in your paper as they
will be removed during generation of camera-ready copies.
}

\cj{
\section{Relationship Between AudioViewer and the Audio-to-Scene, Audio-to-Text Methods}
\hr{We would like to highlight that we do not attempt to compete with but to complement existing high-level audio translation methods.} Ours addresses a scenario that they cannot handle.
The goal of mapping sound to a scene, such as an airplaine or car for their respective engine noise, is to generate the possible corresponding environment image related to the sound information. This does not address the direct translation of sound signals \hr{that is desired for learning to speak and other tasks requiring low-level feedback.} 
\hr{In a similar vein,} mapping speech to text could be applied to adult who want to understand a conversation, but not to children who learn to read only at age 6-8, and not for learning pronounciation in general as no tonal feedback is given.
Learning to speak requires a low-level mapping, like the frequency visualization currently applied and the method we developed. We believe it is particularly important as early childhood learning is hampered by hearing deficits. The related work section highlights the advantages and disadvantages. %
\hr{We cannot use these methods as baselines as they do not apply to our setting, e.g., to distinguish individual phones.} 
The goal of our approach is to learn a phoneme-level mapping between sound and visual signals. So we choose the most closely related low-level mapping methods including audio to lip motion and audio to    
Mel Spectrogram methods as our baselines.    
}

\begin{figure*}[htp]
	\centering
	\includegraphics[width=0.98\textwidth,trim=0 10.3cm 0 0,clip]{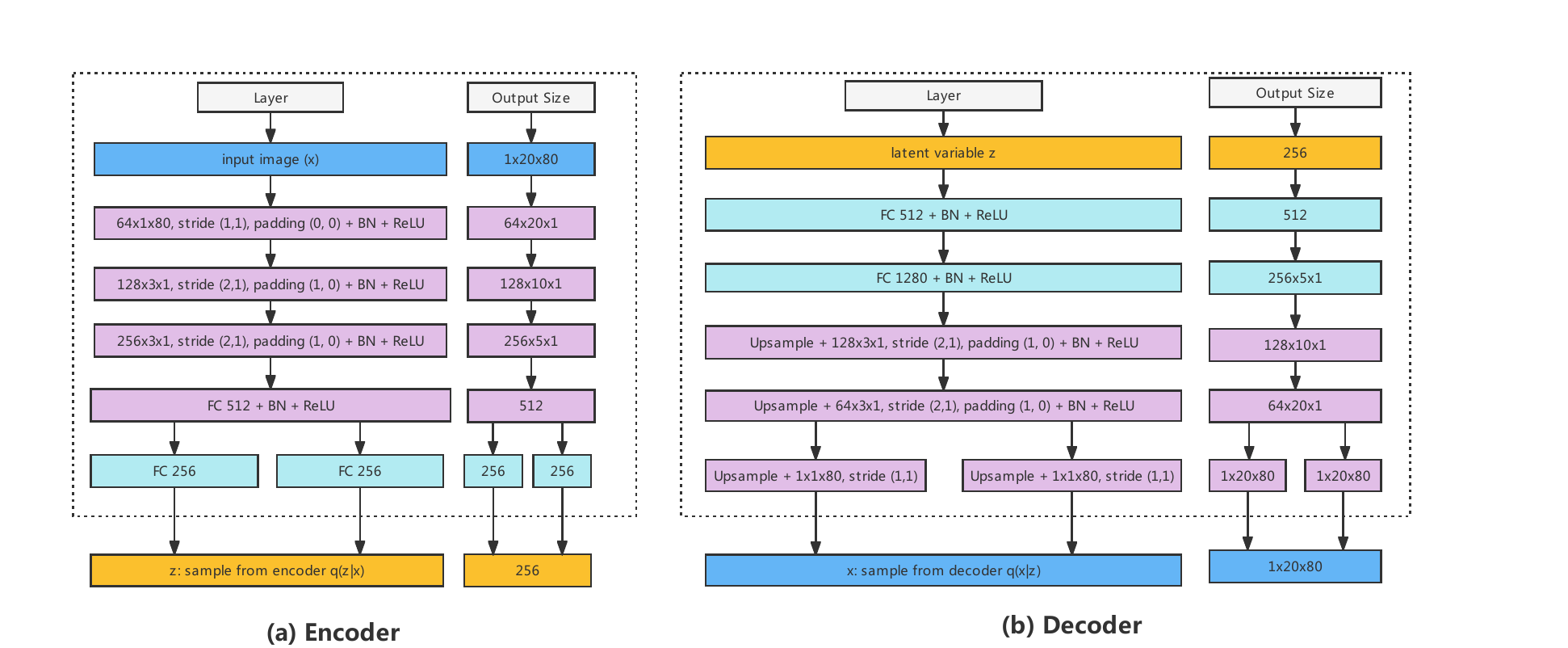}
	\caption{\cj{\textbf{AudioVAE Framework.} 
	(a) and (b) illustrate, respectively, the encoder and decoder parts of the audio model.}}
	\label{fig:net_audiovae}
\end{figure*}

\begin{figure*}[htp]
	\centering
	\includegraphics[width=0.98\textwidth,trim=0 9cm 0 0,clip]{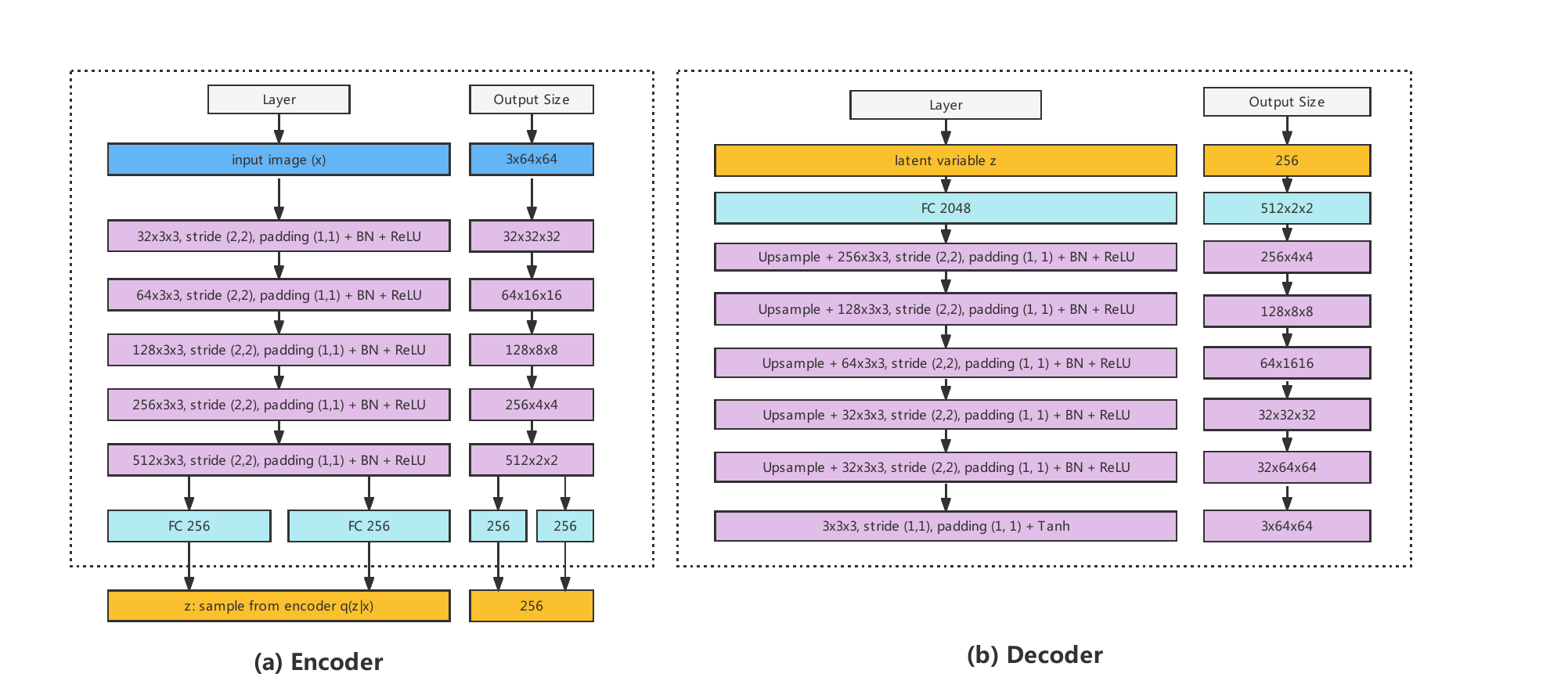}
	\caption{\cj{\textbf{VisualVAE Framework.} 
	(a) and (b) illustrate, respectively, the encoder and decoder parts of the video model.}}
	\label{fig:net_visualvae}
\end{figure*}

\begin{figure*}[htp]
	\centering
	\includegraphics[width=0.98\textwidth,trim=0 14.5cm 0 0,clip]{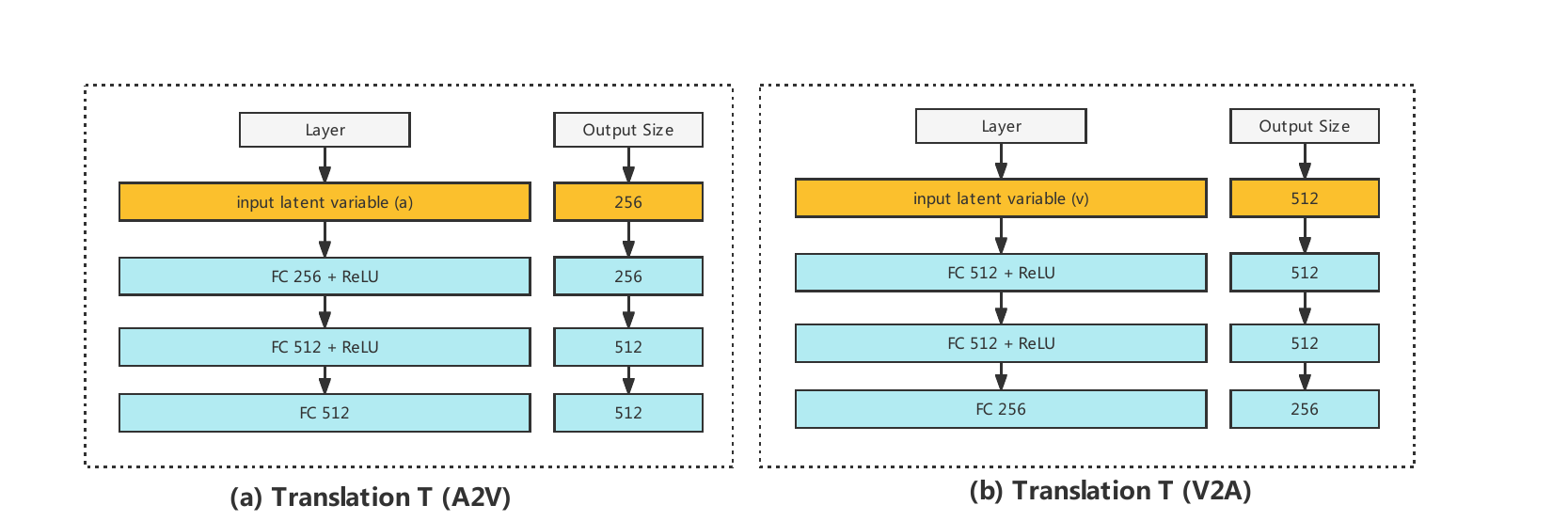}
	\caption{\textbf{Translation Networks.} 
	(a) and (b) illustrate the translation networks mapping audio signal to visual signal, and vice verse.}
	\label{fig:net_T}
\end{figure*}

\cj{
\section{Implementation Details}
The architecture of audio VAE is shown in Figure \ref{fig:net_audiovae}.  We train the audio modules for 300 epochs with batch size 128 and initial learning rate $10^{−3}$. 
We train the low resolution CelebA visual models for 38 epochs and MNIST visual models for 24 epochs with batch size 144 and fixed learning rate $0.005$. Figure \ref{fig:net_visualvae} shows its architecture. 
For the visual model with high resolution, we use the pre-trained Soft-IntroVAE model ~\cite{daniel2021soft} provided by the authors. 
We use two different strategies to link the audio and visual latent spaces. When mapping the audio inputs to low resolution facial and digital images, we fix the audio model and fine tune the visual model on audio and image examples for 10 audio epochs. When linking the audio signals to high resolution facial images, we train the translation module $T$ instead of the entire visual model, whose architecture is shown in Figure \ref{fig:net_T}, with fixed audio model and visual model on audio examples for 10 audio epochs. 
For optimization, we use Adam \cite{kingma2014adam} with parameters $\beta_1 = 0.95$, $\beta_2 = 0.999$, $\epsilon = 10^{-8}$.  

}

\section{Results on Environment Sounds}

With the style disentangling training of our \ours{} method, the audio model is more specific to human speech than general sounds. In the following, we test the generalization capability of our method (trained on the speach TIMIT dataset \cite{garofolo1993timit}) on the ESC-50 environment sound dataset \cite{piczak2015esc}.

\begin{table}[b]
\caption{\textbf{Information throughput on environment sounds,} showing that the reconstruction error increases when evaluating on a test set that contains sounds vastly different from the training set (speech vs. environment sounds).}
\label{tab:cycle_esc}
\resizebox{1\linewidth}{!}{
\begin{tabular}{llr}
\toprule
Audio models                            & Visual models             &  SNR(dB) \\
\midrule
\multirow{1}{*}{Audio PCA}              
                                        & Visual PCA                     & 17.23   \\
                                         \cline{1-3}
\multirow{4}{*}{SpeechVAE  }    %
                                                      & DFC-VAE on CelebA   & 1.03      \\
                                                      & DFC-VAE on MNIST    & 2.22      \\ 

                                             & DFC-VAE on CelebA (refined w/ $\mathcal{L}_{cycle}$)  & \textbf{2.83}   \\
                                             & DFC-VAE on MNIST (refiend w/ $\mathcal{L}_{cycle}$)   & 0.76  \\
                                                      \cline{1-3}
                                                    
\multirow{4}{*}{\begin{tabular}[c]{@{}l@{}}SpeechVAE w/ $\mathcal{L}_{p, \log MSE},$ \\ $\mathcal{L}_{rr}$, dim=256  \end{tabular}} 
                                                      & DFC-VAE on CelebA   & 0.46     \\
                                                      & DFC-VAE on MNIST    & 0.46    \\ 

                                     & DFC-VAE on CelebA (refined w/ $\mathcal{L}_{cycle}$)  & 1.26   \\
                                     & DFC-VAE on MNIST (refiend w/ $\mathcal{L}_{cycle}$)   & \textbf{1.68} \\
                                   
\bottomrule
\end{tabular}}
\end{table}

\begin{table}[b]
\caption{\textbf{Audio VAE mel spectrum reconstruction.} The average SNR of autoencoding and decoding mel spectrograms on ESC-50 shows a significant reconstruction loss. The average speed and acceleration between the latent vector (dim = 128) of neighbouring frames ($\Delta t = 0.04 \mathrm{s}$) confirms the experiments in the main document, that smoothness comes at the cost of lower reconstruction accuracy.}
\label{tab:audio_only_esc}
\resizebox{1\linewidth}{!}{ 
\begin{tabular}{lrrr}
\toprule
Audio models   &   SNR (dB) & Velocity ($\mathrm{s}^{-1}$)  &  Acc.($\mathrm{s}^{-2}$)   \\ 
\midrule
Audio PCA  &   \textbf{17.23}   &    \textbf{170.13}    &   \textbf{6960.80}  \\
SpeechVAE\cite{hsu2017learning}  &  10.17    &  172.57    &   7331.77 \\
\cline{1-4}
{SpeechVAE w/ $\mathcal{L}_{p, \log MSE}$ } & \textbf{8.92} & 58.78 & 1859.95 \\
{SpeechVAE w/ $\mathcal{L}_{rr}$ }  & 2.98 & 40.54 & 1580.86 \\
{SpeechVAE w/ $\mathcal{L}_{p, \log MSE},  \mathcal{L}_{rr}$ } 
                                    & 2.19 & \textbf{30.66} & \textbf{909.39} \\
{SpeechVAE w/ $\mathcal{L}_{p, \log MSE},  \mathcal{L}_{rr}$, dim=256 }
                                     & 2.51 & 33.88 & 1037.97 \\
\bottomrule
\end{tabular}}%
\end{table}

Table~\ref{tab:cycle_esc} shows the reconstruction accuracy when going via the audio and video VAEs (see Information Throughput section in the main document). The SNR for reconstructed Mel spectrum is generally lower than the speech dataset, which is expected since it was trained on the latter. The analysis of the reconstruction ability of the audio VAE in isolation (without going through the video VAE) reported in Table~\ref{tab:audio_only_esc} shows that a large fraction of this loss of accuracy stems from the learning of speech specific features of the audio VAE. 
Moreover, with a recombined reconstruction loss term on the human speech dataset, the model was fitted to speech features and tended to loss high pitch information.
Still, according to the face visualization of the content encoding as we showed in the supplemental video, our \ours{} can generate consistent visualization to given environment sounds. \looseness=-1

\section{Disentangling content and style}

We construct a SpeechVAE that disentangles the style (speaker identity) content (phonemes) in the latent encodings, i.e., the latent encoding $\mathbf{z} = [z_1,\cdots,z_d]^T \in \mathcal{R}^d$ can be separated as a style part $\mathbf{z}_s = [z_1,\cdots,z_{m}]^T$ and a content part $\mathbf{z}_c = [z_{m+1},\cdots,z_{d}]^T$, where $d$ is the whole audio latent space dimension and $m$ in the audio style latent space dimension. \looseness=-1

\begin{figure}[t]
	\centering
 	\includegraphics[width=1\linewidth, trim=0 11.1cm 22cm 0.25cm,clip	]{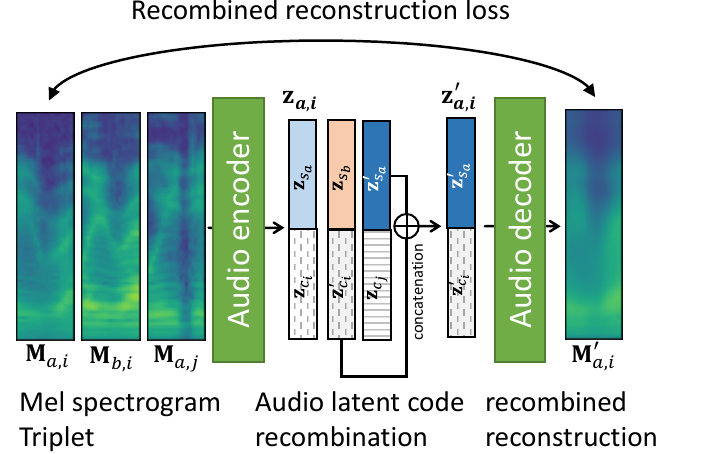}
	\caption{\textbf{Disentanglement}, by mixing content encodings from two speakers saying the same phones ($\vz_{c_i}$ with $\vz'_{c_i}$) and style encodings from the same speaker saying different words ($\vz_{s_a}$ with $\vz'_{s_a}$).}
	\label{fig:disentanglement}
\end{figure}

We use an audio dataset with phone and speaker ID annotation. However, this still requires to disentangle the audio signal into style and content codes, which we obtain 
similarly to \cite{qian2019autovc} by mixing embeddings from different speakers.
 Figure~\ref{fig:disentanglement} gives an overview. At training time, we feed triplets of mel spectrogram segments 
$\mathbf{T}_{a,b,i,j} = \{\mathbf{M}_{a,i},\mathbf{M}_{b,i},\mathbf{M}_{a,j}\}$, where $\mathbf{M}_{a,i}$ and $\mathbf{M}_{b,i}$ are the same phoneme sequence $p_i$ spoken by different speakers $s_a$ and $s_b$ respectively, and $\mathbf{M}_{a,j}$ shares the speaker $s_a$ with the first segment but a different phoneme sequence. Each element of the input triplet is encoded individually by $E_A$, forming latent triplet $\{\mathbf{z}_{a,i},\mathbf{z}_{b,i},\mathbf{z}_{a,j}\} = \{[\mathbf{z}_{s_a},\mathbf{z}_{c_i}]^T, [\mathbf{z}_{s_b},\mathbf{z}_{c_i}']^T,[\mathbf{z}_{s_a}',\mathbf{z}_{c_j}]^T\}$, instead of reconstructing the inputs from the corresponding latent encodings in an autoencoder, we reconstructed the first sample $ \mathbf{M}_{a,i}$ from a recombined latent encoding of the other two, $\mathbf{z}_{a,i}'=[\mathbf{z}_{s_a}',\mathbf{z}_{c_i}']^T $. Formally, we replaced the reconstruction loss term in the VAE objective by a recombined reconstruction loss term,
\begin{equation}
    {\mathcal{L}_{rr}(\mathbf{T}_{a,b,i,j}) = \mathbb {E} _{q_{\phi }(\mathbf {\mathbf{z}_{a,i}'} |\mathbf{M}_{b,i},\mathbf{M}_{a,j} )}{\big (}\log p_{\theta }(\mathbf{M}_{a,i} |\mathbf{z}_{a,i}' ){\big )}}.
\end{equation}
This setup forces the model to learn separate encodings for the style and phoneme information while not requiring additional loss terms.

Note that we could alternatively enforce $\mathbf{z}_{a,i}$ to be close to $\mathbf{z}_{a,i}'$ without decoding (the unused $\mathbf{z}_{a,i}$ in Figure~\ref{fig:disentanglement}). However, an additional L2 loss on the latent space led to a bias towards zero and lower reconstruction scores than the proposed mixing strategy that works with the original VAE objective.

\begin{figure}[t]
	\centering
	\includegraphics[width=0.98\textwidth,trim=0 0cm 0.0cm 0,clip]{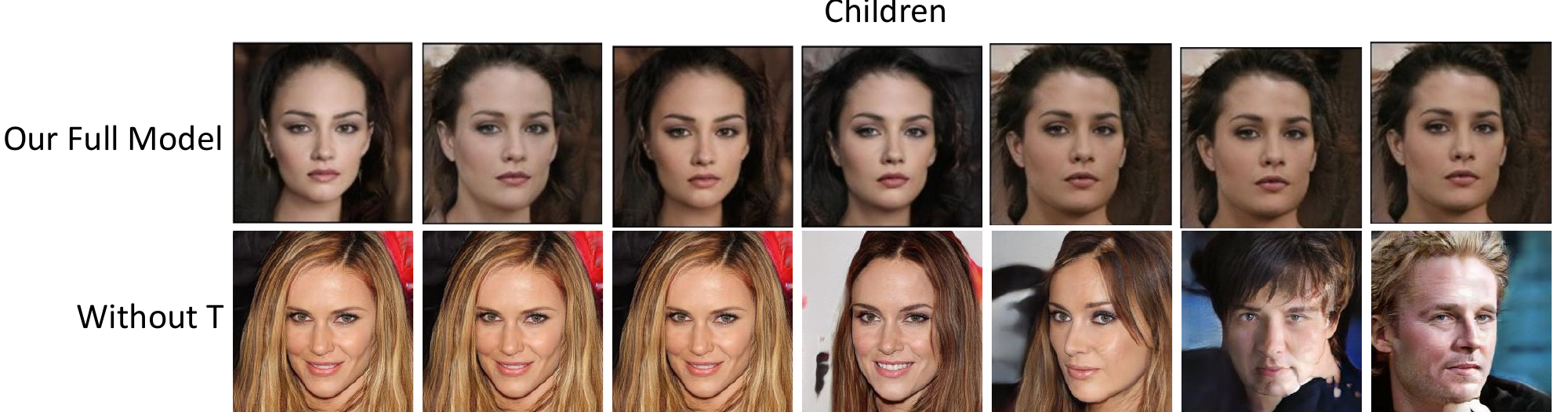}
	\caption{\textbf{The functionality of the translation network $T$.} Here we simply repeat the audio latent codes to match different latent space dimensions and support. Compared with the results of our full model, the smoothness of the generated image sequences is poor.}
	\label{fig:ablationT}
\end{figure}
\section{Ablation study on translation network $T$}
The translation network $T$ is used to build the bridge between audio-vision latent spaces with different dimensions and support. To evaluate the function of $T$, we tried in preliminary experiments to replace $T$ by simply repeating the audio latent code to match latent space dimension. Figure ~\ref{fig:ablationT} shows that the resulting image sequences have 
poor smoothness as their support is not matched. This is confirmed with a low throughput score of $-0.22$, showing that the sound information is lost by mapping to regions that the image decoder does not support.

\comment{
\section{Additional ablation study}

\begin{table}[b]
\caption{\textbf{
Experiments with different cycle loss weight $\lambda_c$.
} Average SNR of cycle-reconstructed amplitude Mel spectrogram on the TIMIT test set.}
\label{tab:cycle}
\resizebox{1\linewidth}{!}{
\begin{tabular}{lllr}
\toprule
Audio models                            & Visual models             & $\lambda_c$ & SNR(dB) \\
\midrule
\multirow{8}{*}{\begin{tabular}[c]{@{}l@{}}SpeechVAE w/ $\mathcal{L}_{p, \log MSE}$ \\ $(\lambda_p=10^3) $ \end{tabular}}
                                   & DFC-VAE on CelebA  & -         & 1.51    \\
                                   & DFC-VAE on MNIST   & -         & 2.16    \\
                                   & DFC-VAE on CelebA (refined w/ $L_c$) & 1        & 4.86     \\
                                   & DFC-VAE on MNIST (refined w/ $L_c$)  & 1        & 4.54     \\ 
                                   & DFC-VAE on CelebA (refined w/ $L_c$) & 10        & \textbf{11.81}  \\
                                   & DFC-VAE on MNIST (refined w/ $L_c$)  & 10        & 3.98  \\ 
                                   & DFC-VAE on CelebA (refined w/ $L_c$) & 100        & 10.51        \\
                                   & DFC-VAE on MNIST (refined w/ $L_c$)  & 100        & 5.22        \\ 
                                   
                                   \cline{1-4}
\multirow{8}{*}{\begin{tabular}[c]{@{}l@{}}SpeechVAE w/ $\mathcal{L}_{p, Q}$ \\ $(\lambda_p=10^3) $ \end{tabular} }
                                   & DFC-VAE on CelebA  & -         & 1.96    \\
                                   & DFC-VAE on MNIST   & -         & 1.87    \\
                                   & DFC-VAE on CelebA (refined w/ $L_c$) & 1        & 5.95     \\
                                   & DFC-VAE on MNIST (refined w/ $L_c$)  & 1        & 7.06     \\ 
                                   & DFC-VAE on CelebA (refined w/ $L_c$) & 10        & \textbf{12.25}  \\
                                   & DFC-VAE on MNIST (refined w/ $L_c$)  & 10        & 3.12  \\ 
                                   & DFC-VAE on CelebA (refined w/ $L_c$) & 100        & 8.71        \\
                                   & DFC-VAE on MNIST (refined w/ $L_c$)  & 100        & 5.26        \\ 
\bottomrule
\end{tabular}}
\end{table}

\begin{table}[b]
\caption{\textbf{Additional ablation study on 
$\lambda_p$}. The average SNR of reconstructed amplitude Mel spectrogram on the TIMIT test set and the average speed and acceleration between the latent vector (dim = 128) of neighbouring frames ($\Delta t = 0.04 \mathrm{s}$).}
\label{tab:audio_only}
\resizebox{1\linewidth}{!}{
\begin{tabular}{lllrrr}
\toprule
Audio models      & $\lambda_p$ &  SNR (dB) & Velocity ($\mathrm{s}^{-1}$)  &  Acceleration ($\mathrm{s}^{-2}$)   \\ 
\midrule
Audio PCA  &   -   &   23.37    &  329.02  &  13395.11 \\
SpeechVAE w/o $\mathcal{L}_{p}$ &   -   &   21.90    &  280.09    &   11648.17 \\
\cline{1-5}
\multirow{3}{*}{SpeechVAE w/ $\mathcal{L}_{p, \log MSE}$ } & 1  & 16.74  & 231.91   & 9490.04  \\
                                                & $10^3$    & \textbf{19.86} & 108.89 & 3052.19 \\
                                                & $10^7$    & 11.43  & 117.43  & 2674.86 \\ 
                                                & $10^9$    & 7.77   & 76.50     & 1735.97   \\ 
                                                \cline{2-5}
\multirow{3}{*}{SpeechVAE w/ $\mathcal{L}_{p, Q}$ } & 1 & 18.83  & 234.19  & 9595.15  \\
                                                & $10^3$    & 18.44  & 119.94  & 3133.26 \\
                                                & $10^7$    & 11.35  & 123.78  & 2476.44 \\ 
                                                & $10^9$    & 9.00  & 119.60   & 2405.18 \\ 
                                                \cline{2-5}
\multirow{3}{*}{SpeechVAE w/ $\mathcal{L}_{p, MSE}$ } & 
                                            $1$    & 17.89  & 243.55   & 10067.36  \\
                                         & $10^3$    & 18.15  & 226.74  & 9252.76 \\
                                         & $10^7$    & 12.64   & 129.10  & 4192.21 \\ 
                                         & $10^9$    & 10.56   & 84.05  & 2709.96 \\ 
\bottomrule
\end{tabular}}%
\end{table}

\HR{Need to check this one. Is it still up-to-date?}
Due to the low SNR of the disentangled model, here we report the ablation experiments on $\lambda_c$ and $\lambda_p$ without the disentangled training to better illustrate the performance of the two loss terms.

Table \ref{tab:cycle} shows the reconstruction error on TIMIT, when reconstructing the mel spectrum via the audio and video VAEs (see Information Throughput section in the main document). We compare versions where the visual VAEs are trained independently or refined with the audio ones using different weights $\lambda_c$ for the cycle loss. Irrespective of the visual domain (CelebA or MNIST), $\lambda_c=10$ works best together with both temporal losses and is selected for the disentangled version. Moreover, as we have expected, the higher cycled reconstruction SNR indicates a higher capacity of information of face visualization than the figure ones.

Table \ref{tab:audio_only} shows the reconstruction SNR (encoding followed by decoding with the audio VAE) on the test set with models trained with different loss functions, as in the main document but with a larger number of tested weights $\lambda_p$. 
The SpeechVAE w/ $\mathcal{L}_{p, log MSE}$ with $\lambda_p=10^3$ is the best performing model that was selected for disentangled model.
}

\section{Human study I - Discriminating Sounds}

The study was conducted with 22, 14, 15, 12, 14, 14, 14 participants for the CelebA-HQ-content, CelebA-content, CelebA-style, CelebA-combined, MNIST-content, MakeItTalk \cite{zhou2020makeittalk} and mel spectrogram (MEL) questionnaires, respectively. 
Each version of the questionnaire asked the same set of 29 questions with randomized ordering of answers within each question. It took participants between 10-15 minutes to complete the each questionnaire. The questionnaire asked participant to perform two possible tasks: matching and grouping visualizations. The format of the questionnaire is outlined in Table~\ref{tab:user_study_format}. The questions tested for two factors: sound content, sounds that share the same phoneme sequences, and sound style, sounds produced by speakers of the same sex or speaker dialect. 
\HR{In total, we tested 100 different sounds and words.}
This purely visual comparison allows us analyze different aspects of the translation task individually.

\begin{figure}[t]
	\centering
	\includegraphics[width=0.45\textwidth]{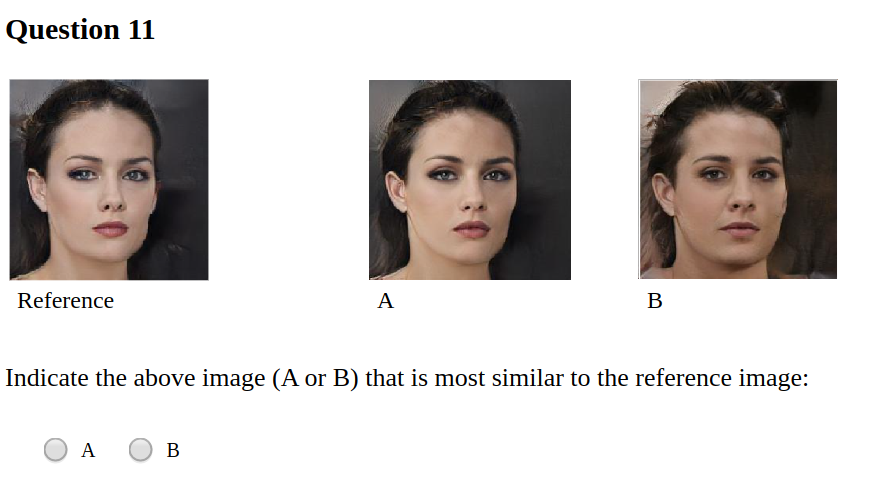}
		\includegraphics[width=0.45\textwidth]{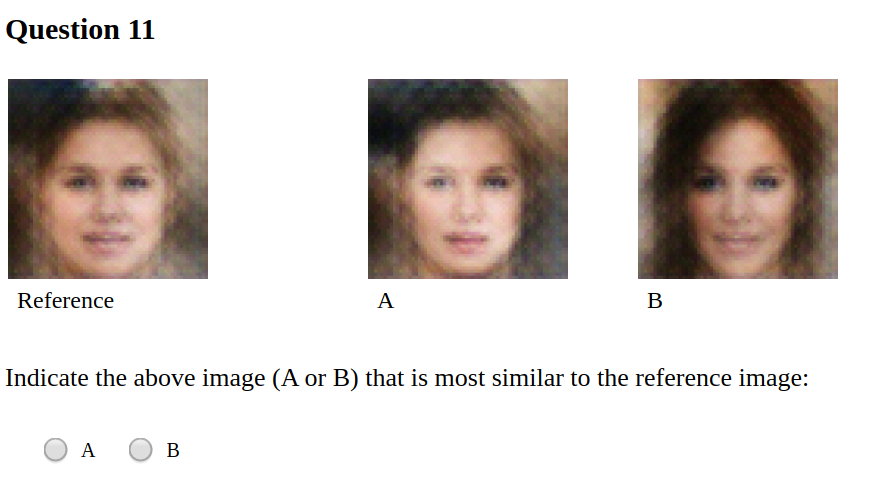}
		\quad\quad
	\includegraphics[width=0.45\textwidth]{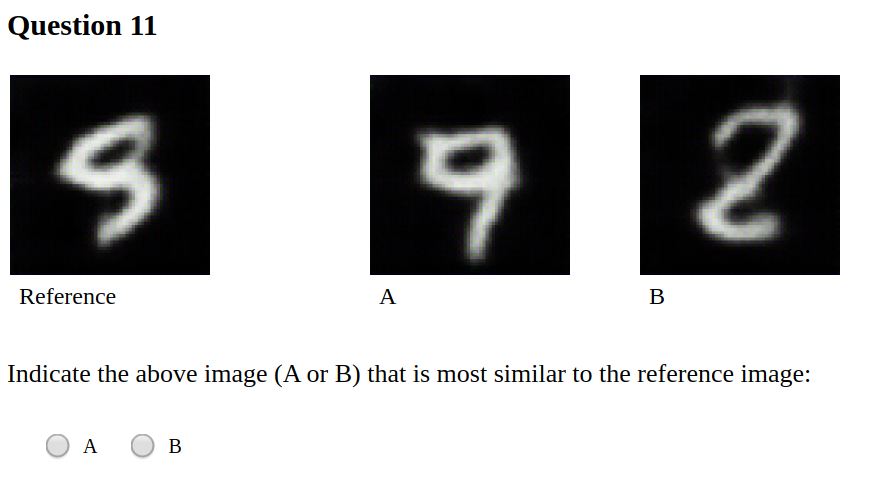}
		\includegraphics[width=0.45\textwidth]{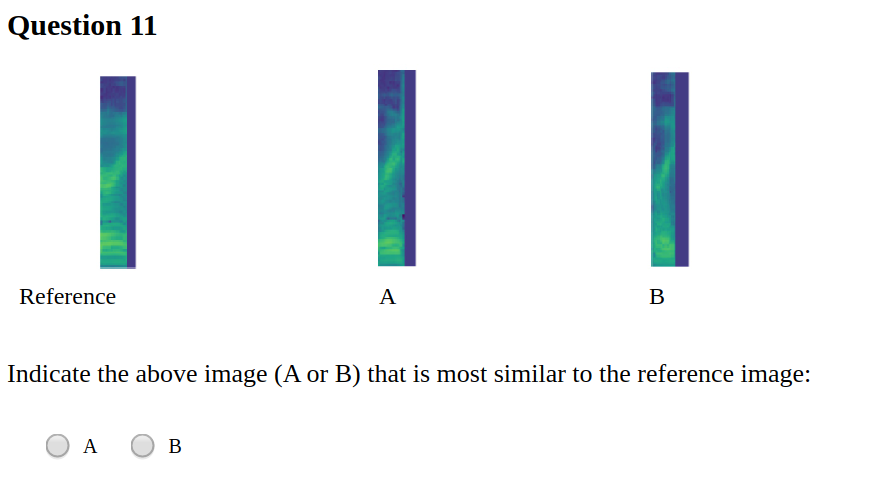}
	\caption{\textbf{Matching Example.} Examples from the CelebA-HQ-content, CelebA-combined, MNIST-content, MEL questionnaire of a matching question asked to participants in the user study. }
	\label{fig:match_example}
\end{figure}
\begin{figure}[t]
	\centering
	\includegraphics[width=0.45\textwidth]{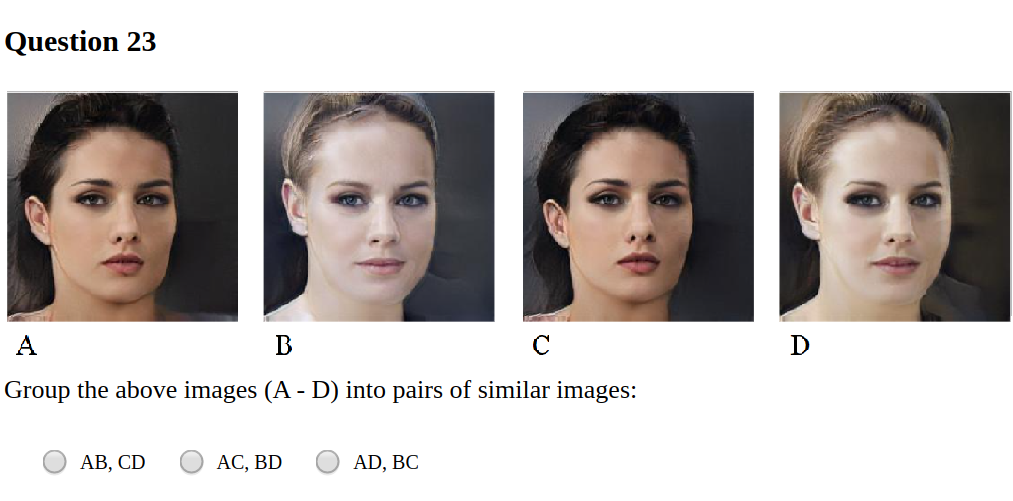}
	\includegraphics[width=0.45\textwidth]{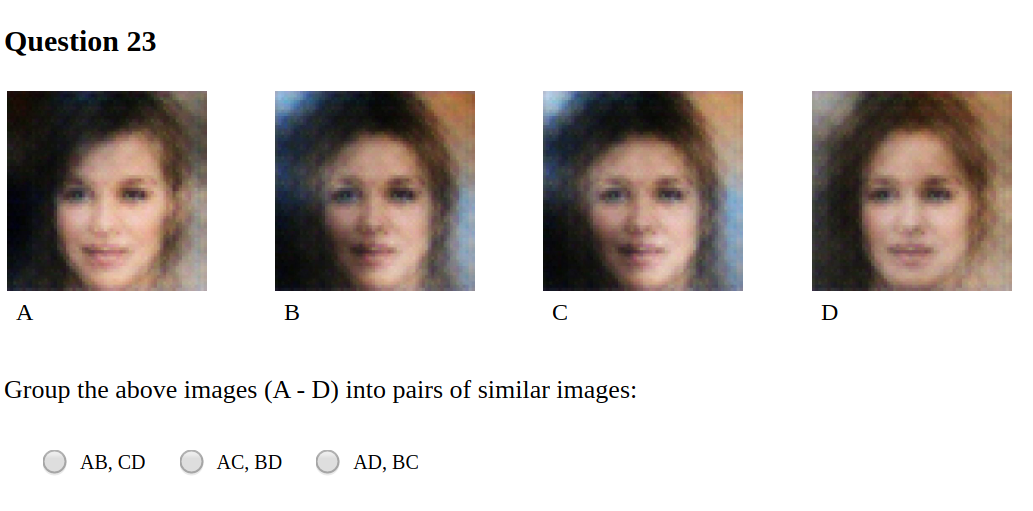}
	\includegraphics[width=0.45\textwidth]{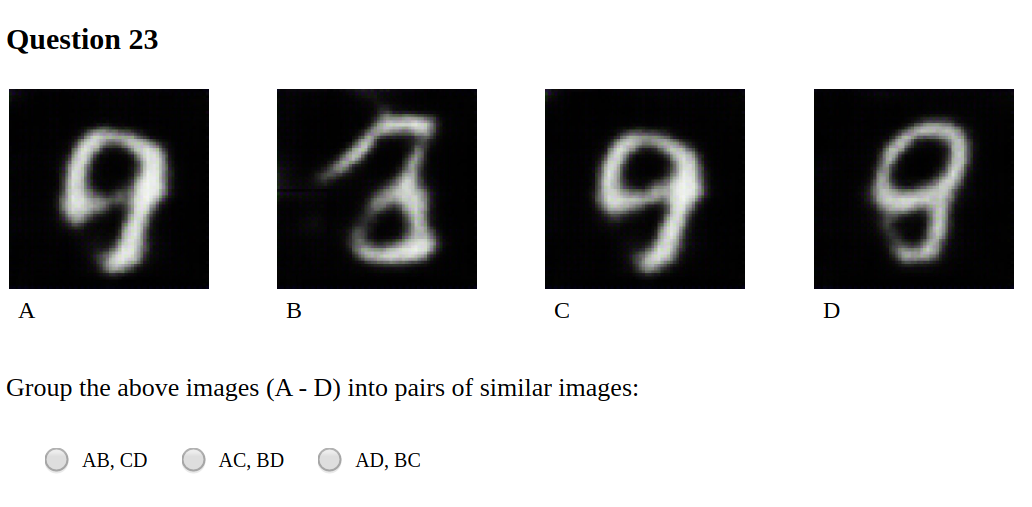}
	\includegraphics[width=0.45\textwidth]{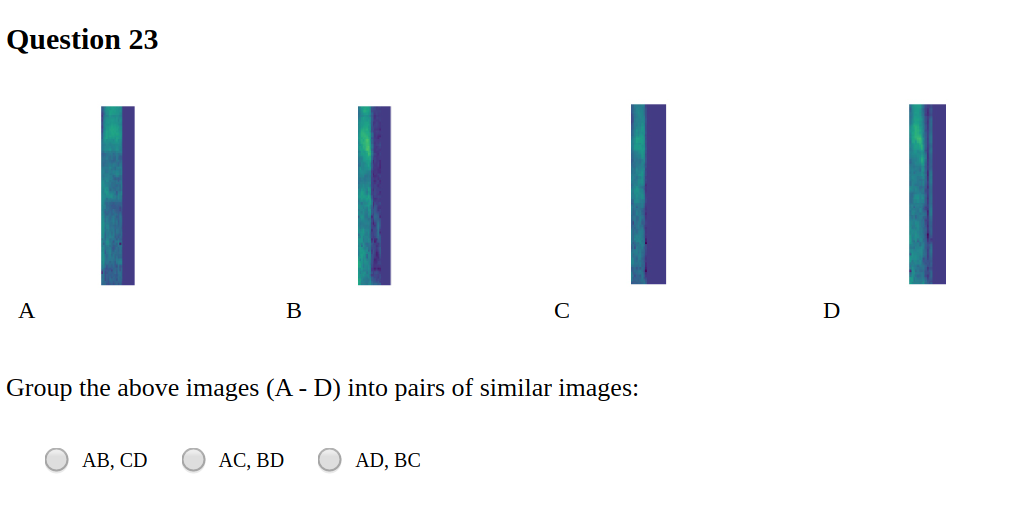}
	\caption{\textbf{Grouping Example.} Examples from the CelebA-HQ-content, CelebA-combined, MNIST-content and MEL questionnaire of a grouping question asked to participants in the user study.}
	\label{fig:group_example}
\end{figure}
\begin{table}[]
    \caption{\textbf{A breakdown of the questionnaire format} by sound type and tested factors, listing their frequency of occurrence.}
    \centering
\resizebox{1\linewidth}{!}{
    \begin{tabular}{lllc}
    \toprule
    Question type & Sound type & Tested factor & Questions \\
    \midrule
        \multirow{6}{4em}{Matching questions} &\multirow{3}{4em}{Phone-pairs} & Content & 3 \\
        & & Style (sex) & 3 \\
        & & Style (dialect) & 2 \\
        \cline{2-4}
        & \multirow{3}{4em}{Words} & Content & 3 \\
        & & Style (sex) & 3 \\ 
        & & Style (dialect) & 2 \\
    \midrule
        \multirow{6}{4em}{Grouping questions} &\multirow{2}{4em}{Phone-pairs} & Content + style (dialect) & 2 \\
        & & Content + style (dialect + sex) & 2 \\
        \cline{2-4}
        & \multirow{5}{4em}{Words} & Content & 3 \\
        & & Content + style (dialect) & 2 \\ 
        & & Content + style (sex) & 1 \\
        & & Content + style (dialect + sex) & 1 \\
        & & Content (similar sounding words) & 2 \\
    \midrule 
    \textbf{Total} & &  & 29 \\
    \bottomrule
    \end{tabular}
    }
    \label{tab:user_study_format}
\end{table}

\paragraph{Matching questions} Matching questions asked the participants to choose which of two possible visualizations which is most visually similar to a given reference visual. Figure~\ref{fig:match_example} shows examples of matching questions. Matching questions were used to assess the viability for users to distinguish between the same sounds produced by speakers possessing different speaker traits as well as determining whether structural similarities in the underlying audio translated into similarities in the visualization. In particular, the questionnaire contained 6 questions for evaluating the ability to distinguish between sound content, which compared visualizations of sounds of different phoneme sequences (3 for phoneme-pairs and 3 for words). Phone-pairs are short in length and therefore the corresponding visualisation was a single frame image, whereas visualisations of words were videos. In order to evaluate the ability to distinguish between sound style, 6  questions compared visualizations of the same phoneme sequence between male and female speakers and 4 questions for distinguishing between speakers of different dialects. In total there were 16 matching questions. Since each question has two options, the expected mean accuracy for random guessing is 50\%.

\parag{Grouping questions.} Grouping questions asked the participants to group 4 visualizations into two pairs of similar visualizations. Figure~\ref{fig:group_example} shows examples of grouping questions. Grouping questions were used to assess the degree to which visualizations of different words are distinguishable and visualizations of the same word are similar. In particular, the study required users to group visualizations of two pairs of sounds, whereby different pairs are sound clips with shared factors of the same sound content or same sound style. In total, the human study consisted of 4 grouping questions based on phone-pairs and 9 grouping questions based on words. Since there are three possible options, the expected mean accuracy for random guessing is 33.3\%.

\parag{Results.}
For each of the models, we tested for sound content: phoneme sequences, and sound style: speaker dialect and speaker sound. We generated the mean accuracy and standard deviation for each tested factor and each question sub type. Table~\ref{tab:user_study_results} extends the results shown in the main document by comparing entangled and disentangled representations. The results of the disentangled models with CelebA visualizations (CelebA-disentangled) is aggregated by taking the results of CelebA-content on the questions which tested for sound content and the results of CelebA-style on questions which tested for sound style.

\begin{table}[t!]
    \caption{\textbf{User study results.} Values indicate mean accuracy and standard error for distinguishing between visualizations of the tested factor across participants as a percentage. The disentangled representation clearly outperforms the combined baseline.}
    \centering
\resizebox{1\linewidth}{!}{
    \begin{tabular}{lccc}
    \toprule
   
    Tested Factor & PCA-Baseline  & CelebA-disentangled & CelebA-combined \\
    \midrule
        
        Content  & 38.4   & $\textbf{85.0} \pm \textbf{\phantom{0}1.8}$  & $72.8 \pm 2.9$ \\
        \midrule
        Style (dialect)  & 50.0  & $\textbf{56.7} \pm \textbf{5.7}$ & $39.6 \pm 7.2$ \\
        \midrule
        Style (sex)  & 43.3 & $\textbf{78.0} \pm \textbf{2.8}$ & $43.3 \pm \phantom{0}2.6$ \\
        \bottomrule
    \end{tabular}
    }
    \label{tab:user_study_results}
\end{table}

\comment{
\begin{figure}[t]
	\centering
	\includegraphics[width=0.5\textwidth,trim=0 12.5cm 25cm 0,clip]{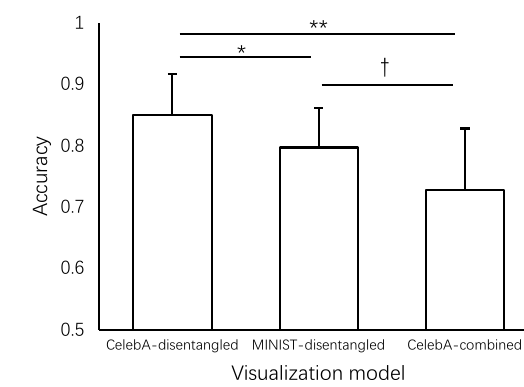}
	\caption{\textbf{Model Comparison on content questions.} The mean user accuracy and standard deviation for distinguishing phoneme sequences is compared between the models, annotated with the significance level (\dag{} for $0.05<p<0.1$, * for $p<0.05$, ** for $p<0.01$). The mean accuracy for random guessing is 0.384.}
	\label{fig:phoneme_comparison}
\end{figure}
}

\begin{figure}[t]
	\centering
	\includegraphics[width=0.5\textwidth]{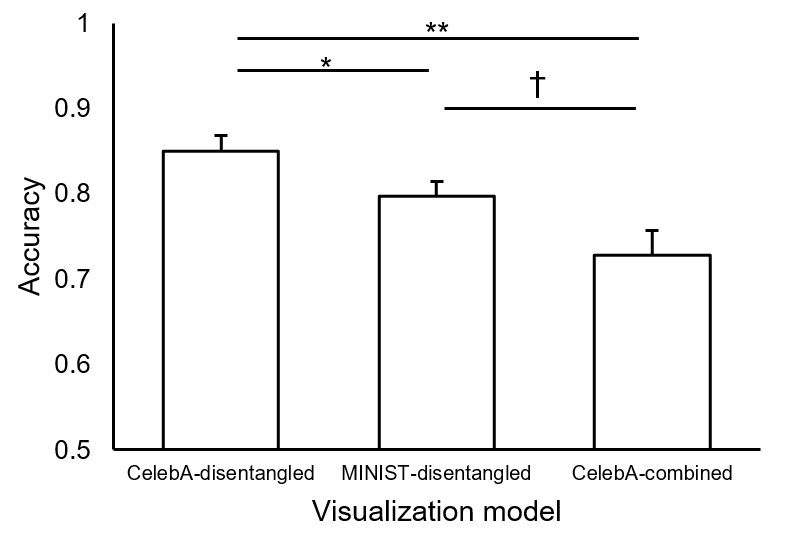}
	\caption{\textbf{Model Comparison on content questions.} The mean user accuracy and standard error for distinguishing phoneme sequences is compared between the models, annotated with the significance level (\dag{} for $0.05<p<0.1$, * for $p<0.05$, ** for $p<0.01$). The mean accuracy for random guessing is 0.384.}
	\label{fig:phoneme_comparison}
\end{figure}
\comment{
\begin{figure}[t]
	\centering
	\includegraphics[width=0.235\textwidth,trim=0 12.5cm 27cm 0,clip]{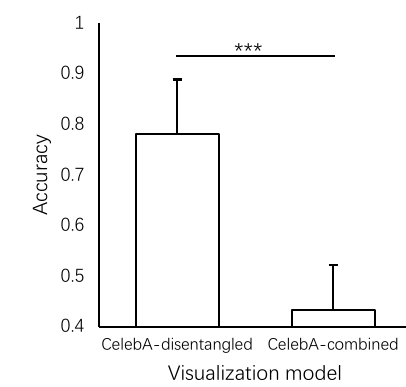}
	\includegraphics[width=0.235\textwidth,trim=0 12.5cm 27cm 0,clip]{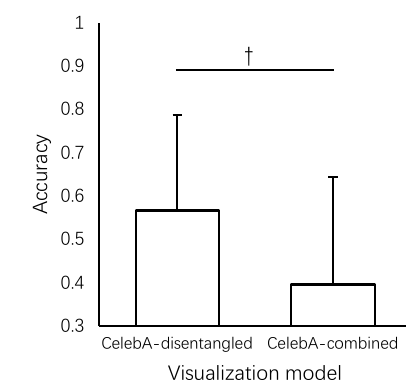}
	\caption{\textbf{Model Comparison on style questions.} The mean user accuracy and standard deviation for distinguishing speaker sex (left) and dialect(right) is compared between the models, annotated with the significance level (\dag{} for $0.05<p<0.1$, *** for $p<0.001$). The mean accuracies for random guessing are 0.43 and 0.5 respectively.}
	\label{fig:style_comparison}
\end{figure}
}

\begin{figure}[t]
	\centering
	\includegraphics[width=0.45\textwidth]{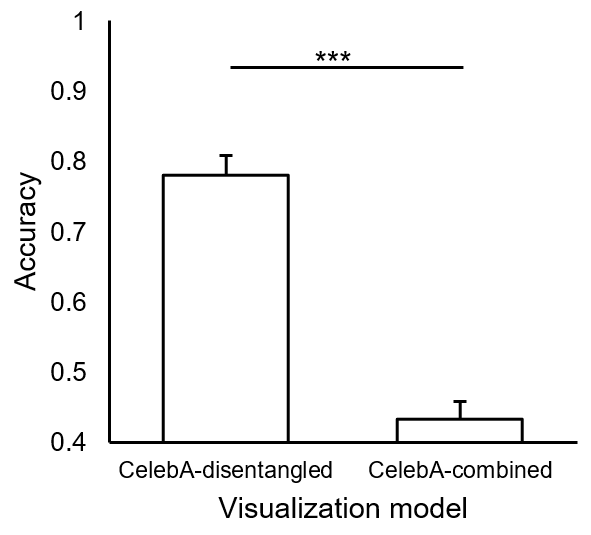}
	\includegraphics[width=0.45\textwidth]{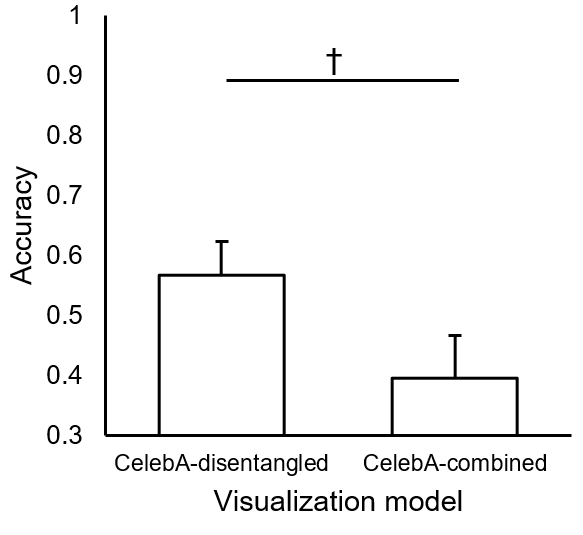}
	\caption{\textbf{Model Comparison on style questions.} The mean user accuracy and standard error for distinguishing speaker sex (left) and dialect(right) is compared between the models, annotated with the significance level (\dag{} for $0.05<p<0.1$, *** for $p<0.001$). The mean accuracies for random guessing are 0.43 and 0.5 respectively.}
	\label{fig:style_comparison}
\end{figure}

We analyze the significance of our improvements by reporting accuracy, standard error, and using the student-t test. For MEL the accuracy is $66.2 \pm 1.3$  with ($p=7\mathrm{e}{-12}$) and for MakeItTalk \cite{zhou2020makeittalk} the accuracy is is $47.7 \pm 2.7$  with ($p=6\mathrm{e}{-12}$). 
Figure~\ref{fig:phoneme_comparison} shows the significance of our other models, it illustrates that users achieve the overall accuracy on the CelebA-disentangled model with $85.0 \pm 1.8$\% (significant with $p < 0.05$) for distinguishing between visualizations of different content. The MNIST-content model has the highest accuracy for distinguishing between different phone pairs with $91.8 \pm 2.5$\%, although not significantly higher than the CelebA-disentangled one with ($p>0.05$), but has a much lower accuracy for distinguishing between different words, suggesting that the MNIST visualizations may be better suited for representing shorter sounds. The CelebA-disentangled model outperforms the CelebA-combined model for distinguishing between speakers of different sex with $78.0 \pm 2.8$\% (significant with $p < 0.001$) and between speakers of different dialects with $56.7 \pm 5.7$\% (marginally significant with $0.05<p<0.10$) as shown in Figure~\ref{fig:style_comparison}. 
The task of distinguishing between different speakers of different dialects is much more difficult than distinguishing between phoneme sequences since there are 8 categories of dialects in the dataset and differences in dialects are much more subtle and can contain often contain overlaps. 
Significance comparing model means were calculated using a two-sample two-tailed t-test with unequal variance and without any outlier rejection.

\section{Human study II - Learning Sounds}
\begin{figure}[t]
	\centering
	\includegraphics[width=1\textwidth]{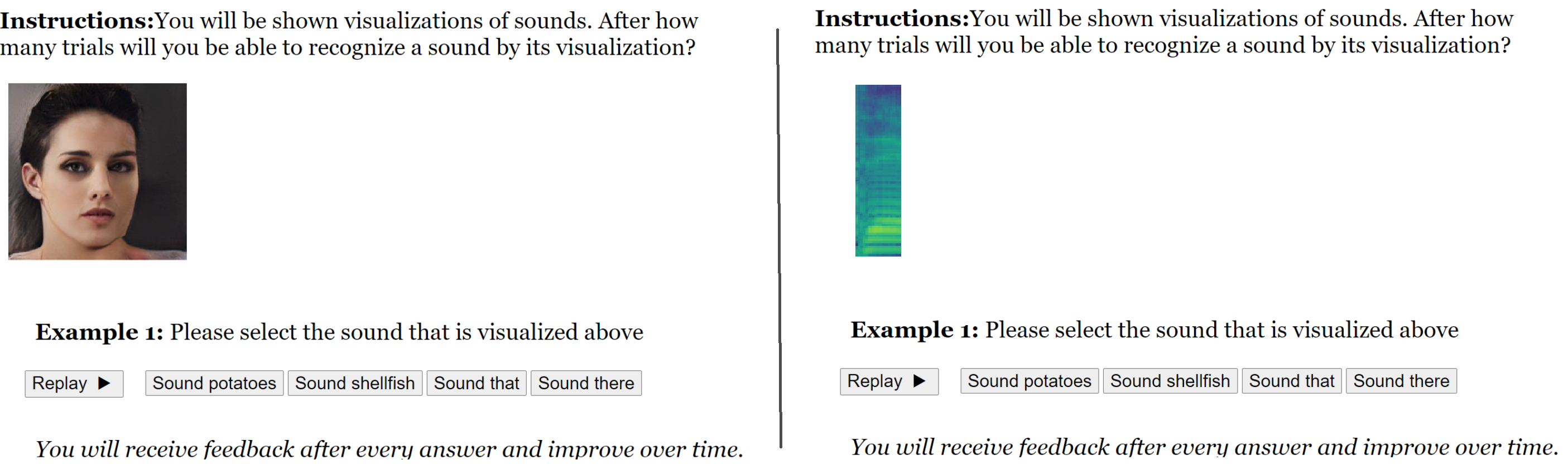}
	\caption{\textbf{Learning study examples.} A video generated by the disentangled content model (left)/MEL (right) is shown to evaluate the participants' capability in learning to recognize sounds from visual contexts. Each video corresponds to one variant of four word labels.}
	\label{fig:learning_study_example}
\end{figure}

\begin{figure}[t]
	\centering
	\includegraphics[width=0.98\linewidth,trim=0.5cm 0cm 0.2cm 0.5cm,clip]{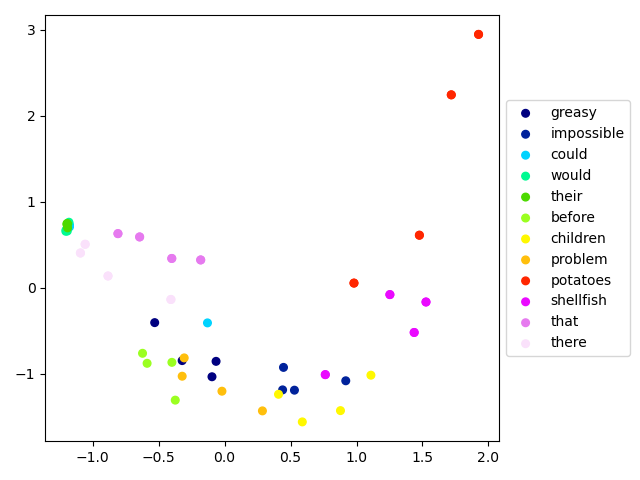}
	\caption{\textbf{Sound embedding distance} visualized by the latent embedding of each word's speech projected to 2D. The same color points correspond to different sounds of the same word.%
	}
	\label{fig:user_study2_MDS}
\end{figure}
\cj{
In the second study, we evaluate whether participants can learn to recognize sounds from our visualizations. 
\hr{Conceivable large scale studies would require a large amount of time to train users, which would take months and and be very expensive when scaling to a representative group size. It is hence not effective for comparing our algorithm variations and unsuitable for establishing a benchmark for future methods.} 
To more effectively evaluate methods, we set human studies in a simple but representative environment that can quickly be repeated with a new cohort of participants to compare the latest methods. 
Specifically, as described in the main paper, we design a small speech dataset containing 16 words, 4 words with different meanings spoken by different people. Among them, 'that' and 'there' are pronounced with similar length and phonemes \HR{(phone editing distance of one or two, depending on the pronunciation)}, while 'potatoes' and 'shellfish' are pronounced in similar length but contain completely different phonemes (editing distance 8 or more). \hr{Note that we chose word pairs that have a similar length as words of different length would make it easy to distinguish simply by the length of the generated video.}

\hr{To further analyze the distance between the chosen words,} we visualize the latent distance between sounds in Figure~\ref{fig:user_study2_MDS}. We randomly select 12 words, each word is spoken by 4 different people. We first encode them to the latent space. Then we expand the latent codes of all words to the same dimension by filling in one for shorter ones. Finally, we embed them into two dimensions using PCA to preserve distances in the original high-dimensional space. This embedding shows that 'there' and 'that' is very close in the embedding while 'patatoes' and 'shellfish' are separated by other words (light green). Please note that this is a non-linear embedding from high-dimensional space that best approximates distances but contains some deformation. Moreover, due to the concatenation and one padding, a slight differences in duration or speed would lead to quite different word embeddings. However, we did not see another way of visualizing such word embedding in a 2D space.
}

To showcase the robustness to different speakers, we included words spoken by people with two different dialects. To ensure a fair comparison to using the MEL specturm baseline, we decided to only test on words spoken by the same gender (male speakers). Otherwise, the MEL spectrum of female speakers would look very different from male speakers due to their higher pitch. This would lead to slower learning on MEL and results would hence only support the better gender normalization of our method (as validated in Study I) instead of validating better learnability. Words are given in random order to participants in order to avoid any bias towards the visualizations that appear first.

\hr{We considered testing the learning of sentences or phones (cf. Study I on distinguishing but not learning). However, entire sentences are longer and contain more information than a single word. Hence, it would be easier to distinguish them. Moreover, single phones map only to a single frame and hence would not test the continuous translation into a video. Since} our approach is a low-level sound visualization method, it is more challenging and suitable to select words than either sentences or single phones for evaluating learnability.

Examples of user learning are shown in Figure~\ref{fig:learning_study_example}. 
The learnability of our method is evaluated by comparing the tracking of the changes in the accuracy curves between our model and MEL.

\parag{Results.}
We recruited 9 participants for human study II.
It took participants between 10-15 minutes to complete each variant. \cj{During the learning period, samples are shown in a random order to avoid bias to the ordering.} The learning curves are reported in the main document. 
The accuracy after 16 rounds of learning is for Ours $87.0\%$ vs. MEL $57.8\%$, a significant improvement with ($p = 0.016$).
We conclude that compared to spectogram representations, our mapping to images of faces or digits is more natural and easier for people to distinguish and match. Whether remaining ambiguities could be overcome by longer learning sessions and how the learning can be further facilitated remains an open question for future work. The main paper discusses the results in more detail.

\cj{}

{\small
\bibliographystyle{ieee_fullname}
\bibliography{tex/references}
}